\ifdefined\pdfminorversion\pdfminorversion=7\fi
\documentclass[preprint,12pt]{elsarticle}

\usepackage{amssymb}
\usepackage{amsmath}

\usepackage{subcaption}

\usepackage[normalem]{ulem}

\usepackage{microtype}
\usepackage[hidelinks,hypertexnames=false]{hyperref}
\biboptions{sort&compress}

\journal{Journal of Computational Physics}

\begin{document}

\begin{frontmatter}



\title{GUERNICA: A continuum kinetic code for neutral dynamics in fusion plasmas}


\author{Jack Gabriel\corref{first}} 
\ead{jgabriel@wm.edu}
\cortext[first]{Corresponding author.}
\author{Saskia Mordijck}

\affiliation{organization={Department of Physics, William \& Mary},
            city={Williamsburg},
            state={VA 23185}, 
            country={USA}}

\begin{abstract}
We introduce GUERNICA, a GPU-accelerated continuum kinetic code for atomic hydrogen transport in fusion edge plasmas. It solves the Boltzmann equation with charge exchange, ionization, recombination and elastic neutral--neutral collisions. GUERNICA discretizes two--dimensional configuration space with the discontinuous Galerkin method on unstructured meshes and velocity space with the discrete velocity method in three dimensions, enabling direct simulation of the neutral distribution function without statistical noise. Unlike Monte Carlo methods, which introduce statistical noise and become computationally expensive in strongly collisional regimes, GUERNICA provides a deterministic alternative. An FFT-based evaluation of the full charge exchange operator reduces its computational complexity from \(O(N_v^2)\) to \(O(N_v\log N_v)\), enabling computationally practical simulations in three-dimensional velocity space. We verify GUERNICA against analytic tests and validate it against one-dimensional Monte Carlo results from DEGAS2. The numerical tests demonstrate the expected convergence orders and discrete conservation properties, while the DEGAS2 comparisons show agreement in neutral density and temperature to within 1.6\%. Performance studies confirm the expected scaling of the FFT-based charge exchange operator and demonstrate efficient execution on a single GPU. Finally, we demonstrate GUERNICA in two dimensions using a static plasma background from a MAST-U SOLPS-ITER simulation, with \(8.6\times10^8\) phase-space degrees of freedom on a single GPU,  illustrating the method's ability to resolve neutral dynamics in experimentally relevant divertor conditions.

\end{abstract}


\end{frontmatter}



\section{Introduction}
\label{sec:Into}
Neutral atoms and molecules play an essential role in the edge and divertor regions of magnetically confined fusion plasmas. In these low-temperature boundary regions, neutrals are sourced primarily by wall recycling and gas puffing, and they participate in a wide range of processes including charge exchange, ionization, recombination and elastic scattering \cite{mordijck2020, stangeby2000}. Through these interactions, neutrals strongly influence plasma fueling, parallel and cross-field momentum balance, impurity generation and, critically, the reduction of divertor heat loads in detached operating regimes \cite{zholobenko2021, krasheninnikov2016}. The ability of neutrals to penetrate deep into the scrape-off layer (SOL) and modify the distribution of particle and energy fluxes makes accurate neutral transport modeling an indispensable component of predictive simulations for present devices as well as for ITER and future power-plant--scale reactors \cite{kukushkin2005}.

Kinetic Monte Carlo (MC) solvers, including EIRENE \cite{reiter2005} and DEGAS2 \cite{stotler1994}, dominate neutral modeling in fusion research. They are embedded in widely used plasma edge packages including SOLPS-ITER \cite{wiesen2015}, SOLEDGE2D-EIRENE \cite{bufferand2013} and EMC3-EIRENE \cite{frerichs2019}. MC approaches offer excellent geometric flexibility and can naturally incorporate the complex atomic and molecular physics required in reactor-relevant regimes. However, these advantages come with two major drawbacks: statistical noise and collisional-regime sensitivity. As neutral--ion mean free paths become short, typical in high-density divertor plasmas or partially detached conditions, the number of collisions per particle trajectory increases dramatically. MC methods then struggle to converge, often requiring extremely large particle counts and long averaging times to reduce variance. Moreover, the presence of MC noise complicates steady-state convergence when these neutral solvers are coupled iteratively to plasma fluid or gyrokinetic codes, an issue that becomes increasingly severe as recycling coefficients approach unity in large devices \cite{joseph2017}.

These limitations have motivated the development of alternative approaches. Fluid-neutral models provide dramatic speedups in highly collisional regimes but rely on closures that break down when neutral mean free paths become large \cite{moscheni2022}. Hybrid techniques like spatial decomposition, micro-macro formulations and asymptotic-preserving MC schemes aim to bridge kinetic and fluid limits and have demonstrated success in specific settings \cite{van_uytven2022, borodin2022}. Deterministic continuum kinetic methods, including discrete ordinates and discrete velocity methods (DVM), have seen recent application in rarefied gas flows and divertor vacuum pumping (e.g., DIVGAS \cite{tantos2020}), and continuum neutral models have been implemented in gyrokinetic or drift-reduced frameworks such as Gkeyll \cite{bernard2022} and GBS \cite{giacomin2022}. Yet, to our knowledge, no continuum kinetic neutral code has combined the full charge exchange operator, unstructured two-dimensional geometry and GPU acceleration for the edge/divertor environment of fusion plasmas.

To address this gap, we introduce GUERNICA (Galerkin-based Unstructured Element Representation of Neutral Interactions and Collisions Algorithm), a new deterministic kinetic neutral transport code. GUERNICA solves the Boltzmann equation for atomic hydrogen neutrals using the discrete velocity method (DVM) in velocity space and the discontinuous Galerkin (DG) method in configuration space, enabling direct resolution of the full neutral distribution function without statistical noise. The model includes the full charge exchange operator, a linear ionization operator, a recombination source and a Bhatnagar--Gross--Krook (BGK) model for neutral--neutral collisions, enabling modeling over a wide range of collisional regimes.

GUERNICA is built on the MFEM finite-element library \cite{anderson2021}, leveraging matrix-free evaluation, GPU-friendly kernels and efficient data layouts to achieve high throughput even for large 2X3V simulations. Fast Fourier Transform (FFT)--based acceleration reduces the computational complexity of the charge exchange operator from $\mathcal{O}(N_v^2)$ to $\mathcal{O}(N_v\log N_v)$, allowing high-resolution velocity grids that are impractical for direct quadrature.

The remainder of the paper is organized as follows. Section \ref{sec:GovEq} presents the governing equations and collision operators, including charge exchange, ionization and BGK neutral--neutral collisions as well as the recombination source. Section \ref{sec:Num} details the numerical discretization using DG in configuration space and DVM in velocity space, as well as the implementation of collision operators, time stepping and boundary conditions. Section \ref{sec:VV} verifies the individual operators, transport and boundary conditions, characterizes the stability and computational cost of the solver and validates GUERNICA against DEGAS2. Section \ref{sec:MAST} applies GUERNICA to a two-dimensional MAST-U edge geometry using a SOLPS-ITER background plasma. Section \ref{sec:Conc} summarizes the main results and discusses future extensions, including molecules, multi-GPU scaling and coupling to plasma solvers. Overall, this paper shows that GUERNICA is a noise-free continuum alternative to Monte Carlo for atomic hydrogen transport and lays the groundwork for its use in high-density, high-collisionality divertor conditions, where Monte Carlo methods are most expensive.

\section{Governing equations}
\label{sec:GovEq}

We describe atomic hydrogen neutrals in the edge and divertor regions of a magnetically confined plasma by the phase-space distribution function $f_n(\mathbf{x},\mathbf{v},t)$, which satisfies the Boltzmann equation,
\begin{equation}
\label{eq:Gov}
    \frac{\partial f_n}{\partial t} + \mathbf{v}\cdot\nabla_\mathbf{x} f_n = C[f_n]+S_{\text{rec}}+S_n.
\end{equation}
Here $\mathbf{x}\in\mathbb{R}^d$ denotes configuration space, with $d=1 \text{ or } 2$ here, and $\mathbf{v}\in\mathbb{R}^3$ is the neutral velocity. The collision operator $C[f_n]$ collects the processes that depend on the neutral distribution, $S_{rec}$ is the volumetric source of neutrals from electron--ion radiative recombination and $S_n$ represents externally prescribed volumetric sources of neutrals such as gas puffing or recycling sources.

Macroscopic neutral quantities are obtained as velocity moments of $f_n$. The neutral density, bulk velocity and temperature are defined as,
\begin{equation*}
\begin{split}
    n_n(\mathbf{x},t) &= \int f_n(\mathbf{x},\mathbf{v},t) d\mathbf{v}, \\
    \mathbf{u}_n(\mathbf{x},t) &= \frac{1}{n_n} \int \mathbf{v}f_n d\mathbf{v}, \\
    T_n(\mathbf{x},t) &= \frac{m_n}{3n_n} \int |\mathbf{v}-\mathbf{u}_n|^2f_n d\mathbf{v},
\end{split}
\end{equation*}
where $m_n$ is the neutral particle mass. We treat the plasma quantities, namely the electron density, $n_e$, ion density, $n_i$, ion flow velocity, $\mathbf{u}_i$, electron temperature, $T_e$, and ion temperature, $T_i$, as externally prescribed fields obtained from a background plasma simulation, and for simplicity assume Maxwellian ions and electrons when evaluating their interactions with neutrals.

\subsection{Collision operators and recombination source}
\label{subsec:ColGov}

GUERNICA includes four atomic processes. Electron-impact ionization removes neutrals through collisions with electrons, and charge exchange redistributes neutral momentum and energy through collisions with ions. Neutral--neutral elastic scattering relaxes the neutral distribution toward local thermodynamic equilibrium, and electron--ion radiative recombination creates neutrals from the plasma. The first three depend on the neutral distribution function and together form the total collision operator,
\begin{equation*}
\label{eq:collisions}
C[f_n] = C_{iz}[f_n] + C_{cx}[f_n] + C_{bgk}[f_n],
\end{equation*}
where $C_{iz}, C_{cx}$ and $C_{bgk}$ correspond to electron-impact ionization, charge exchange and neutral--neutral elastic collisions modeled using the BGK operator \cite{bhatnagar1954}. Recombination involves only the prescribed background plasma and is independent of $f_n$, so it enters Eq.~\ref{eq:Gov} as the separate source $S_{rec}$ rather than as part of $C[f_n]$. All four are represented using effective cross sections or rate coefficients, following standard kinetic neutral modeling practice, and are described in turn below.

\noindent \textbf{Ionization}

Electron-impact ionization removes neutral atoms and is represented by the loss term,
\begin{equation*}
\label{eq:izFull}
C_{iz}[f_n](\mathbf{v}) = - f_n(\mathbf{v}) \int f_e(\mathbf{v}') \sigma_{iz}(|\mathbf{v} - \mathbf{v}'|) |\mathbf{v} - \mathbf{v}'| d\mathbf{v}' ,
\end{equation*}
where $f_e$ is the electron distribution function, $\sigma_{iz}$ is the ionization cross section, $\mathbf{v}$ is the neutral velocity and $\mathbf{v}'$ is the electron velocity. Since electron thermal velocities are much larger than neutral velocities in typical edge plasmas, the relative speed is approximated as
$|\mathbf{v} - \mathbf{v}'| \approx |\mathbf{v}'|$. Under this assumption, the ionization operator reduces to,
\begin{equation}
\label{eq:izMath}
C_{iz}[f_n](\mathbf{v}) = - \nu_{iz}\, f_n(\mathbf{v}),
\end{equation}
with ionization frequency,
\begin{equation*}
\label{eq:izNu}
\nu_{iz} = n_e \langle \sigma_{iz} v_e \rangle(T_e) ,
\end{equation*}
where $\langle \sigma_{iz} v_e \rangle$ is the Maxwellian-averaged ionization rate coefficient evaluated at the local electron temperature.

\noindent \textbf{Charge exchange}
\nopagebreak

Charge exchange between neutrals and ions is represented by a symmetric gain--loss operator,
\begin{align}
\label{eq:CX_full}
C_{cx}[f_n](\mathbf{v}) &= - f_n(\mathbf{v}) \int f_i(\mathbf{v}')
\sigma_{cx}(|\mathbf{v} - \mathbf{v}'|) |\mathbf{v} - \mathbf{v}'| d\mathbf{v}' \nonumber \\
&\quad + f_i(\mathbf{v}) \int f_n(\mathbf{v}') \sigma_{cx}(|\mathbf{v} - \mathbf{v}'|) |\mathbf{v} - \mathbf{v}'| d\mathbf{v}' ,
\end{align}
where $f_i$ is the ion distribution function and $\sigma_{cx}$ is the charge exchange cross section. In both terms, $\mathbf{v}$ is the velocity at which the operator is evaluated and $\mathbf{v}'$ is the integration variable running over the velocities of the collision partner. The first term is the loss of a neutral with velocity $\mathbf{v}$ through charge exchange with an ion with velocity $\mathbf{v}'$. Conversely, the second term is the gain of a neutral with velocity $\mathbf{v}$ created when an ion with velocity $\mathbf{v}$ undergoes charge exchange with a neutral with velocity $\mathbf{v}'$; the collision exchanges the identities of the two particles while leaving their velocities unchanged.

Defining the charge exchange rate kernel,
\begin{equation}
\label{eq:CX_kernel}
K_{cx}(|\mathbf{v} - \mathbf{v}'|) = \sigma_{cx}(|\mathbf{v} - \mathbf{v}'|)\, |\mathbf{v} - \mathbf{v}'|,
\end{equation}
the charge exchange operator can be interpreted as a convolution in velocity space between the neutral and ion distributions. Equation \ref{eq:CX_full} can then be written as,
\begin{equation}
\label{eq:CX_FFT}
    C_{cx}[f_n](\mathbf{v}) = -f_n(\mathbf{v})(K_{cx}*f_i)(\mathbf{v}) + f_i(\mathbf{v})(K_{cx}*f_n)(\mathbf{v}),
\end{equation}
where,
\begin{equation*}
    (K_{cx}*f)(\mathbf{v}) = \int f(\mathbf{v}') K_{cx}(|\mathbf{v} - \mathbf{v}'|) d\mathbf{v}'.
\end{equation*}
GUERNICA exploits this convolution structure in the numerical implementation of Sec. \ref{subsec:ColNum}.

\noindent \textbf{Neutral--neutral collisions}

We model elastic collisions between neutrals with the BGK approximation to the full Boltzmann operator, 
\begin{equation}
\label{eq:bgkMath}
C_{bgk}[f_n](\mathbf{v}) = - \nu_{nn} \big( f_n(\mathbf{v}) - f_{eq}(\mathbf{v}) \big),
\end{equation}
where $\nu_{nn}$ is an effective neutral--neutral collision frequency \cite{bhatnagar1954}, prescribed here as a constant. In the continuous case, the equilibrium distribution $f_{eq}$ is a Maxwellian constructed from the neutral moments as,
\begin{equation*}
\label{eq:Maxwellian}
f_{eq}(\mathbf{v}) = \frac{n_n}{(2\pi T_n / m_n)^{3/2}} \exp\!\left(- \frac{m_n |\mathbf{v} - \mathbf{u}_n|^2}{2 T_n} \right).
\end{equation*}
This operator conserves neutral mass, momentum and energy at the continuous level. The preservation of these invariants in the discrete case depends on the numerical realization of velocity space and is discussed in Sec. \ref{subsec:ColNum}.

\noindent \textbf{Recombination}
\nopagebreak

Electron--ion recombination creates neutral atoms from the background plasma. As with ionization, the relative speed of the reactants is dominated by the electron velocity, so the rate depends only on the electron temperature. Since the electron mass is negligible, the new neutral retains the velocity of the recombining ion. Thus, neutrals are born with the ion distribution,
\begin{equation}
\label{eq:Recomb}
S_{\text{rec}}(\mathbf{v}) = n_e \langle \sigma_{\text{rec}} v_e \rangle(T_e) \, f_i(\mathbf{v}),
\end{equation}
where $\langle \sigma_{\text{rec}} v_e \rangle$ is the Maxwellian-averaged radiative recombination rate coefficient evaluated at the local electron temperature. Three-body recombination, which dominates at low $T_e$ and high $n_e$, is left for future work.

\subsection{Non-dimensionalization}
\label{subsec:Dims}

The quantities in GUERNICA simulations span many orders of magnitude, including densities, thermal velocities, collision rates, source strengths and cross sections. Before discretization, we therefore write the equations in dimensionless form, with each physical quantity $q$ expressed as $q=q_0\tilde{q}$, where $q_0$ is a reference scale and $\tilde{q}$ is the dimensionless counterpart. This scaling improves numerical conditioning, simplifies the implementation of collision and source terms and allows the verification problems of Sec.~\ref{sec:VV} to be posed directly in dimensionless units.

Four reference values are chosen independently: a density $n_0$, length $L_0$, temperature $T_0$ and particle mass $m_0$. These define the reference velocity and time,
\begin{equation*}
v_0 = \sqrt{\frac{e T_0}{m_0}}, \qquad t_0 = \frac{L_0}{v_0},
\end{equation*}
where $T_0$ is expressed in electron volts and $e$ is the elementary charge. Neutrals and ions are the same species here, so $m_0 = m_n = m_i$. The reference scales of the remaining quantities follow from these and are summarized in Table \ref{tab:reference}.

\begin{table}[!htbp]
\centering
\caption{The reference scales used to non-dimensionalize each respective quantity.}
\label{tab:reference}
\begin{tabular}{||cc||}
\hline
Quantity     & Reference Scale \\
\hline
\hline
$\mathbf{x}$ & $L_0$ \\
$t$          & $L_0/v_0$     \\
$\mathbf{v}$, $\mathbf{u}$   &   $v_0$    \\
$n$          &   $n_0$    \\
$T$          &   $T_0$    \\
$f$          &   $f_0 \equiv n_0/v_0^3$    \\
$\sigma$     &   $1/\left(n_0L_0\right)$    \\
$\langle \sigma v \rangle$, $K_{cx}$  & $v_0/\left(n_0L_0\right)$      \\
$\nu$        &   $v_0/L_0$    \\
$C$, $S_n$, $S_{rec}$   &   $f_0v_0/L_0$    \\      
\hline
\end{tabular}
\end{table}

Under these definitions, Eq. \ref{eq:Gov} becomes,
\begin{equation}
\label{eq:Gov_nd}
\frac{\partial \tilde{f}_n}{\partial \tilde{t}} + \tilde{\mathbf{v}}\cdot\nabla_{\tilde{\mathbf{x}}}\tilde{f}_n = \tilde{C}[\tilde{f}_n] + \tilde{S}_{\text{rec}}+\tilde{S}_n.
\end{equation}
The collision operators and recombination source retain their algebraic forms,
\begin{align*}
\tilde{C}_{iz}[\tilde{f}_n](\tilde{\mathbf{v}}) &= - \tilde{n}_e \, \widetilde{\langle \sigma_{iz} v_e \rangle}(\tilde{T}_e) \tilde{f}_n(\tilde{\mathbf{v}}), \\
\tilde{C}_{cx}[\tilde{f}_n](\tilde{\mathbf{v}})
&=
-\tilde{f}_n(\tilde{\mathbf{v}})\,(\tilde{K}_{cx}*\tilde{f}_i)(\tilde{\mathbf{v}})
+\tilde{f}_i(\tilde{\mathbf{v}})\,(\tilde{K}_{cx}*\tilde{f}_n)(\tilde{\mathbf{v}}), \\
\tilde{C}_{bgk}[\tilde{f}_n](\tilde{\mathbf{v}})
&=
-\tilde{\nu}_{nn}\left(\tilde{f}_n(\tilde{\mathbf{v}})-\tilde{f}_{eq}(\tilde{\mathbf{v}})\right), \\
\tilde{S}_{\text{rec}}(\tilde{\mathbf{v}})
&=
\tilde{n}_e \, \widetilde{\langle \sigma_{\text{rec}} v_e \rangle}(\tilde{T}_e)\, \tilde{f}_i(\tilde{\mathbf{v}}).
\end{align*}
The dimensionless neutral moments become,
\begin{align*}
\tilde{n}_n &= \int \tilde{f}_n \, d\tilde{\mathbf{v}}, \\
\tilde{\mathbf{u}}_n &= \frac{1}{\tilde{n}_n}\int \tilde{\mathbf{v}}\,\tilde{f}_n\,d\tilde{\mathbf{v}}, \\
\tilde{T}_n &= \frac{1}{3\tilde{n}_n}\int |\tilde{\mathbf{v}}-\tilde{\mathbf{u}}_n|^2 \tilde{f}_n \, d\tilde{\mathbf{v}},
\end{align*}
where the mass no longer appears because $m_n = m_0$. For notational simplicity, the tildes denoting dimensionless quantities are omitted throughout the remainder of this paper.

\section{Numerical methods}
\label{sec:Num}

\begin{figure}
    \centering
    \includegraphics[width=0.8\linewidth]{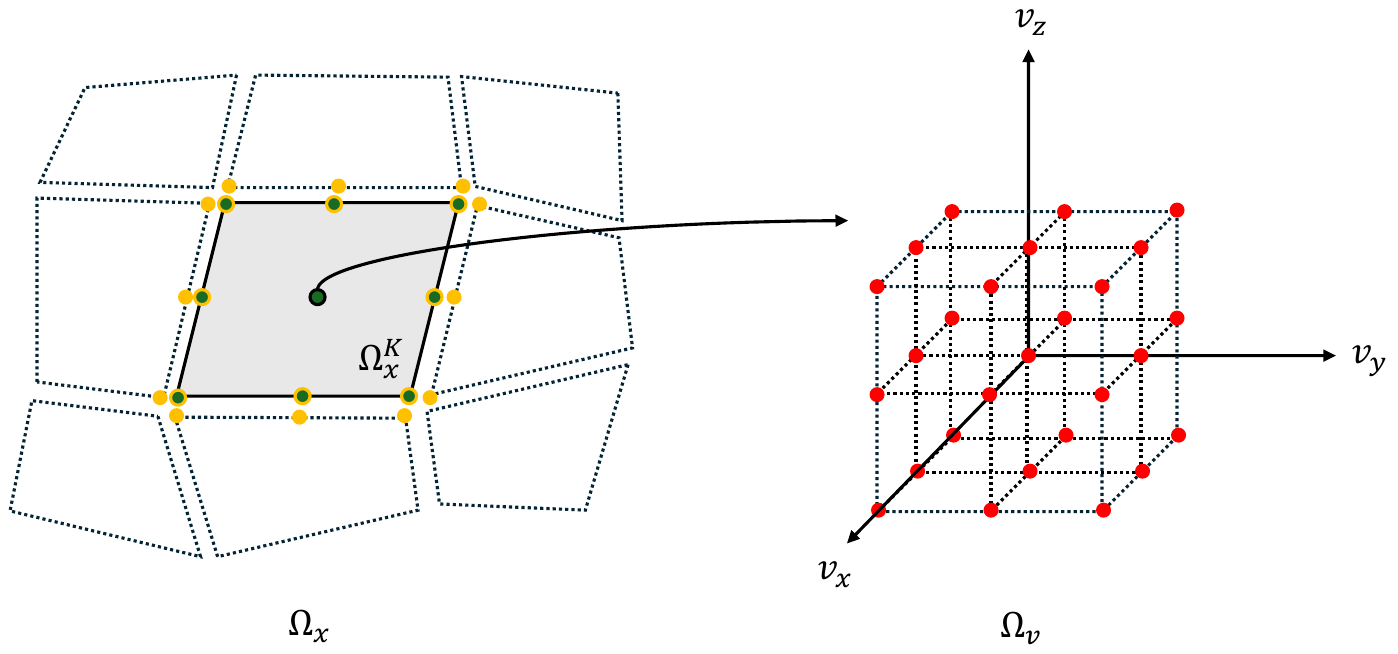}
    \caption{2X3V phase space discretization using an unstructured spatial domain $\Omega_x$ with $\mathbb{P}_2$ quad elements and a velocity domain with $N_v = 27$. Circles represent the spatial nodes (green), flux boundary nodes (orange) and velocity nodes (red), respectively.}
    \label{fig:discretization}
\end{figure}

This section describes the numerical discretization of the neutral kinetic equation in both velocity space and configuration space. We first discretize velocity space with the DVM, producing transport equations at the discrete velocities, and then discretize these equations in configuration space with the DG method. Figure~\ref{fig:discretization} shows a schematic of the discretization.

The configuration-space discretization is built on the MFEM library \cite{anderson2021}, which provides the mesh representation, finite element spaces, element transformations, quadrature data and basis evaluations used by the DG method described in Sec.~\ref{subsec:DG}. GUERNICA implements the velocity-space discretization, the velocity quadrature and all operator kernels, including the matrix-free DG transport operator, on top of this framework.

\subsection{Discrete velocity method}
\label{subsec:DVM}

The DVM replaces the continuous three-dimensional velocity space $\Omega_v$ with a finite set of discrete velocities. Each discrete velocity defines a separate transport equation, coupled to the others only through the collision operators. We first truncate the continuous velocity space to a bounded Cartesian domain,
\begin{equation*}
\mathcal{V} = [v_x^{\min}, v_x^{\max}] \times [v_y^{\min}, v_y^{\max}] \times [v_z^{\min}, v_z^{\max}],
\end{equation*}
which is then discretized using a uniform tensor-product grid in each velocity direction with $N_v$ total discrete velocities, including both endpoints of each interval. The truncation bounds must be chosen sufficiently large to ensure that the support of the distribution function is well contained within $\mathcal{V}$. In practice, the velocity domain is selected such that the thermal tails of the distribution are negligible at the boundaries. The discrete velocities are denoted by,
\begin{equation*}
\mathbf{v}_j = (v_{x,l}, v_{y,m}, v_{z,n}),
\end{equation*}
with constant spacings $\Delta v_x$, $\Delta v_y$, and $\Delta v_z$. The uniform Cartesian velocity grid enables efficient evaluation of convolution-type collision operators, particularly the charge exchange operator of Sec. \ref{subsec:ColNum}.

The distribution function is represented at the discrete velocities as,
\begin{equation*}
f_n(\mathbf{x}, \mathbf{v}, t) \;\longrightarrow\; f_{n,j}(\mathbf{x}, t) = f_n(\mathbf{x}, \mathbf{v}_j, t).
\end{equation*}
For each discrete velocity $\mathbf{v}_j$, Eq. \ref{eq:Gov_nd} reduces to a scalar transport equation in configuration space,
\begin{equation}
\label{eq:DVM}
\frac{\partial f_{n,j}}{\partial t} + \mathbf{v}_j \cdot \nabla_\mathbf{x} f_{n,j} = Q_j[f_n],
\end{equation}
where $Q_j[f_n]$ denotes the collision and source operators evaluated at $\mathbf{v} = \mathbf{v}_j$. The spatial discretization of the transport term is described in Sec. \ref{subsec:DG}.

GUERNICA approximates the velocity integrals appearing in the macroscopic moments and collision operators with a quadrature on the uniform velocity grid. For a generic velocity integral the discrete approximation takes the form,
\begin{equation*}
\int_{\mathcal{V}} g(\mathbf{v})\, d\mathbf{v} \;\approx\; \sum_j g(\mathbf{v}_j)\, w_j,
\end{equation*}
where the quadrature weights $w_j$ are the tensor product of one-dimensional weights in each velocity direction. For an odd number of points in a direction these are the composite Simpson weights while for an even number of points, GUERNICA uses the alternative extended Simpson rule \cite{press2007}. For a distribution that has decayed at the edges of the velocity domain, the even rule reduces to the trapezoidal rule, which converges exponentially for smooth integrands. Section~\ref{sec:transport} compares the two cases.

The same quadrature gives the neutral moments. GUERNICA evaluates the neutral density, momentum density and total energy density,
\begin{align}
n_n &= \sum_j f_{n,j}\, w_j, \label{eq:discreteDensity} \\
n_n \mathbf{u}_n &= \sum_j \mathbf{v}_j f_{n,j}\, w_j, \label{eq:discreteMomentum} \\
E_n &= \sum_j \tfrac{1}{2} |\mathbf{v}_j|^2 f_{n,j}\, w_j. \label{eq:discreteEnergy}
\end{align}
The bulk velocity $\mathbf{u}_n$ and temperature $T_n$ are then recovered from these quantities as,
\begin{equation*}
\begin{split}
\mathbf{u}_n &= \frac{1}{n_n} \sum_j \mathbf{v}_j f_{n,j}\, w_j, \\ T_n &= \frac{2}{3 n_n} \left(E_n - \tfrac{1}{2} n_n |\mathbf{u}_n|^2 \right).
\end{split}
\end{equation*}

\subsection{Discontinuous Galerkin method}
\label{subsec:DG}

We discretize configuration space with the discontinuous Galerkin method, partitioning the two-dimensional domain $\Omega_x \subset \mathbb{R}^2$ into non-overlapping quadrilateral elements $K$. Within each element, the distribution function corresponding to velocity $\mathbf{v}_j$ is approximated by the finite-dimensional expansion,
\begin{equation*}
f_{n,j}(\mathbf{x}, t)\big|_K = \sum_{l=1}^{N_p} f_{l,j}^K(t)\, \phi_l^K(\mathbf{x}),
\end{equation*}
where $\{\phi_l^K\}_{l=1}^{N_p}$ are element-local basis functions and $N_p$ denotes the number of spatial degrees of freedom per element. The order of the approximation is represented by $\mathbb{P}_p$ where $N_p = (p+1)^2$ for quadrilateral elements. The basis functions are allowed to be discontinuous across element interfaces with no continuity constraints imposed between neighboring elements. GUERNICA uses the Lagrange interpolants of the tensor-product Gauss--Lobatto--Legendre (GLL) points as basis functions, and uses the same GLL points as the quadrature nodes for all volume integrals.

Multiplying Eq.~\ref{eq:DVM} by a test function $\phi_m^K$ and integrating by parts over $K$ gives the weak form,
\begin{equation}
\begin{split}
\label{eq:WeakFull}
\sum_{l=1}^{N_p} M_{ml}^K \frac{d f_{l,j}^K}{dt} - v_{j,x} \sum_{l=1}^{N_p} S_{ml}^{K,x} f_{l,j}^K - v_{j,y} \sum_{l=1}^{N_p} S_{ml}^{K,y} f_{l,j}^K \\
+ \int_{\partial K} \phi_m^K (\mathbf{v}_j \cdot \mathbf{n}) \hat f_{n,j} \, dS = \int_K \phi_m^K Q_j[f_n] \, d\mathbf{x},
\end{split}
\end{equation}
where $\mathbf{n}$ is the outward unit normal on the element boundary, $\hat f_{n,j}$ is the numerical flux and the matrices $M_{ml}^K$, $S_{ml}^{K,x}$ and $S_{ml}^{K,y}$ are defined as,
\begin{align*}
\label{eq:MassStiffness}
M_{ml}^K &= \int_K \phi_m^K \phi_l^K \, d\mathbf{x}, \\
S_{ml}^{K,x} &= \int_K \frac{\partial \phi_m^K}{\partial x} \phi_l^K \, d\mathbf{x}, \\
S_{ml}^{K,y} &= \int_K \frac{\partial \phi_m^K}{\partial y} \phi_l^K \, d\mathbf{x}.
\end{align*}
Because the nodal and quadrature points coincide, $M_{ml}^K$ is diagonal,
\begin{equation*}
\label{eq:DiagMass}
M_{ml}^K = \delta_{ml}\, \omega_l \, \det J(\mathbf{x}_l),
\end{equation*}
where $\omega_l$ is the GLL quadrature weight and $\det J(\mathbf{x}_l)$ is the local Jacobian determinant. This collocation also allows GUERNICA to evaluate Eq.~\ref{eq:WeakFull} in matrix-free form by exploiting the tensor-product structure of the basis on quadrilateral elements \cite{deville2002, orszag1980}.

The numerical flux is specified using an upwind flux based on the sign of the normal velocity component. At an interface between two neighboring elements, the flux is defined as,
\begin{equation*}
\hat f_{n,j} =
\begin{cases}
f_{n,j}^{-}, & \mathbf{v}_j \cdot \mathbf{n} > 0, \\
f_{n,j}^{+}, & \mathbf{v}_j \cdot \mathbf{n} < 0,
\end{cases}
\end{equation*}
where $f_{n,j}^{-}$ and $f_{n,j}^{+}$ denote the interior and exterior traces of the distribution function at the interface. This definition is applied independently for each discrete velocity. Equation~\ref{eq:WeakFull} then yields, for each discrete velocity, a system of ordinary differential equations in time for the coefficients $f_{l,j}^K(t)$, coupled between neighboring elements only through the numerical flux.

The spatial transport operator acts independently on each discrete velocity, while the collision operators act independently at each spatial node. This structure motivates an element-major storage layout in which each element owns one contiguous block of memory holding its coefficients for all $N_v$ discrete velocities. Within that block the nodal values of each discrete velocity are themselves contiguous, so the transport kernels, which sweep over the nodes of one velocity, and the collision kernels, which sweep over the velocities of one node, both access memory with a fixed stride.

\subsection{Boundary conditions}
\label{subsec:BCNum}

GUERNICA imposes boundary conditions weakly through the numerical flux described in Sec.~\ref{subsec:DG}. For each boundary face with outward unit normal $\mathbf{n}$, the upwind flux requires an exterior state $f_{n,j}^{+}$ for velocities satisfying $\mathbf{v}_j\cdot\mathbf{n}<0$. Velocities with $\mathbf{v}_j\cdot\mathbf{n}>0$ are outgoing from the computational domain and are determined entirely by the interior trace $f_{n,j}^{-}$. This paper uses two boundary conditions: periodic boundaries and diffuse Maxwellian wall emission.

For periodic boundaries, the exterior trace on one periodic boundary is obtained from the corresponding interior trace on the paired periodic boundary. If $\mathbf{x}$ is a point on a periodic boundary face and $\mathbf{x}_p$ is the corresponding point on the paired face, then the exterior state is prescribed as
\begin{equation*}
\label{eq:periodic_bc}
f_{n,j}^{+}(\mathbf{x}) = f_{n,j}^{-}(\mathbf{x}_p).
\end{equation*}
The upwind numerical flux is then applied in the same way as on an interior face.

Diffuse wall boundary conditions prescribe neutral emission from material surfaces. For a diffuse wall, the incoming distribution is taken to be a Maxwellian determined by the wall temperature $T_w$ and wall velocity $\mathbf{u}_w$. The wall Maxwellian evaluated at the discrete velocity $\mathbf{v}_j$ is
\begin{equation*}
\label{eq:wall_maxwellian}
M_{w,j} = \frac{1}{(2\pi T_w)^{d_v/2}}\exp\left(-\frac{|\mathbf{v}_j-\mathbf{u}_w|^2}{2T_w}\right),
\end{equation*}
where $d_v$ is the dimension of the velocity space represented in the simulation. The incoming exterior state is then prescribed as
\begin{equation*}
\label{eq:diffuse_bc}
f_{n,j}^{+} = \alpha_w M_{w,j}, \qquad \mathbf{v}_j\cdot\mathbf{n}<0 .
\end{equation*}
Here $\alpha_w$ is the wall-emission amplitude. It is either prescribed directly or, for a recycling wall, set on each boundary face so that the emitted neutral flux equals the ion flux $n_i \max(\mathbf{u}_i\cdot\mathbf{n}, 0)$ carried onto that face by the background plasma, corresponding to a recycling coefficient of one. The half-range flux of $M_{w,j}$ is evaluated with the velocity quadrature of Sec.~\ref{subsec:DVM}, so the discrete emitted flux matches its target exactly. Neutrals that reach a wall leave the domain, and no part of the outgoing neutral flux is re-emitted.

\subsection{Collision operators and recombination source}
\label{subsec:ColNum}

GUERNICA evaluates the collision operators and recombination source of Sec.~\ref{subsec:ColGov} at each spatial node using the velocity quadrature of Sec.~\ref{subsec:DVM}. Together with the prescribed source $S_n$ they make up the right-hand side $Q_j[f_n]$ of Eq.~\ref{eq:WeakFull}, which is local in space. The numerical treatment of each term is described below.

\noindent \textbf{Ionization}

Under the approximation described in Sec. \ref{subsec:ColGov}, electron-impact ionization reduces to a linear loss term of the form given in Eq. \ref{eq:izMath}. In the discrete velocity formulation, the ionization operator is therefore evaluated pointwise in velocity space as,
\begin{equation*}
C_{iz}[f_n](\mathbf{v}_j) = -n_e \langle \sigma_{iz} v_e \rangle(T_e) f_{n,j},
\end{equation*}
where the Maxwellian-averaged rate coefficient $\langle \sigma_{iz} v_e \rangle(T_e)$ can be obtained from atomic databases. GUERNICA currently employs the Voronov fit,
\begin{equation*}
    \langle \sigma_{iz} v_e \rangle = A\frac{1+P\left(E_{iz}/T_e\right)^{1/2}}{X+E_{iz}/T_e}\left(\frac{E_{iz}}{T_e}\right)^Ke^{-E_{iz}/T_e},
\end{equation*}
where $A, P, X$ and $K$ are tabulated for elements up to $Z=28$ and $E_{iz}$ is the ionization energy \cite{voronov1997}. For atomic hydrogen, $A = 2.91\times10^{-14}$ m$^3$/s, $P=0, X=0.232,K=0.39$ and $E_{iz}=13.6$ eV.

\noindent \textbf{Charge exchange}

The charge exchange operator of Eq. \ref{eq:CX_full} couples the neutral and ion distributions through velocity integrals involving the kernel $K_{cx}$ of Eq. \ref{eq:CX_kernel}. The cross section $\sigma_{cx}$ combines the Krsti\'c \& Schultz fit \cite{krstic1998} below a center-of-mass energy of 100~eV with the Janev \& Smith fit \cite{janev1993} above it, as in DEGAS2, with the latter rescaled by a constant so that the cross section is continuous at 100~eV. As shown in Eq. \ref{eq:CX_FFT}, this operator can be written as a sum of convolution terms in velocity space.

In the discrete velocity setting, a direct evaluation of these convolutions would require $\mathcal{O}(N_v^2)$ operations per spatial degree of freedom, which becomes prohibitively expensive for three-dimensional velocity grids. GUERNICA instead exploits the uniform Cartesian velocity grid of Sec.~\ref{subsec:DVM} to evaluate the convolutions with fast Fourier transforms (FFTs). It first multiplies each distribution by the quadrature weights and embeds it in a velocity grid padded to at least twice its size in every direction, which avoids aliasing. The discrete convolution is then computed as a pointwise multiplication in Fourier space,
\begin{equation*}
\widehat{K_{cx} * f} = \hat{K}_{cx} \, \hat{f},
\end{equation*}
followed by an inverse transform back to velocity space. This approach reduces the computational complexity of the charge exchange operator to $\mathcal{O}(N_v \log N_v)$ and is applied independently at each spatial node.

The gain and loss terms are assembled using the resulting convolution fields, yielding a discrete realization of Eq. \ref{eq:CX_FFT} that is consistent with the velocity quadrature used for moment evaluation. The convolution can optionally be evaluated in single precision, covering the kernel, the transforms and the convolved fields, while the neutral distribution and the assembly of the gain and loss terms remain in double precision. Sections~\ref{sec:cost}, \ref{sec:degas2} and \ref{sec:MAST} quantify the effect of this option.

\noindent \textbf{Neutral--neutral collisions}

The BGK operator of Eq.~\ref{eq:bgkMath} conserves mass, momentum and energy at the continuous level. In discrete velocity space, conservation requires that the discrete equilibrium distribution reproduce the discrete moments of $f_n$ given by Eqs.~\ref{eq:discreteDensity}--\ref{eq:discreteEnergy} under the same velocity quadrature used elsewhere in the algorithm. We enforce conservation at the discrete level by constructing an equilibrium distribution $f_{eq,j}$ satisfying the matching conditions,
\begin{align}
\sum_{j=1}^{N_v} f_{eq,j}\, w_j &= n_n, \label{eq:bgk_match_rho}\\
\sum_{j=1}^{N_v} \mathbf{v}_j f_{eq,j}\, w_j &= n_n\boldsymbol{u}_n, \label{eq:bgk_match_mom}\\
\sum_{j=1}^{N_v} |\mathbf{v}_j|^2 f_{eq,j}\, w_j &= 2E_n. \label{eq:bgk_match_E}
\end{align}

Following the approach by Mieussens, $f_{eq}$ is represented in exponential form,
\begin{equation}
\label{eq:bgk_exp}
f_{eq,j} = \exp\!\left(\alpha_0 + \boldsymbol{\alpha}_1 \cdot \mathbf{v}_j + \alpha_2 |\mathbf{v}_j|^2 \right),
\end{equation}
where the unknown coefficients $\alpha_0 \in \mathbb{R}$, $\boldsymbol{\alpha}_1 \in \mathbb{R}^3$ and $\alpha_2 \in \mathbb{R}$ are determined such that Eqs. \ref{eq:bgk_match_rho}--\ref{eq:bgk_match_E} hold \cite{mieussens2000}. Defining the constraint vector,
\begin{equation*}
\mathbf{G}(\boldsymbol{\alpha})=
\begin{bmatrix}
\sum_j f_{eq,j}(\boldsymbol{\alpha}) w_j - n_n \\
\sum_j v_{j,x} f_{eq,j}(\boldsymbol{\alpha}) w_j - (n u_x)_n \\
\sum_j v_{j,y} f_{eq,j}(\boldsymbol{\alpha}) w_j - (n u_y)_n \\
\sum_j v_{j,z} f_{eq,j}(\boldsymbol{\alpha}) w_j - (n u_z)_n \\
\sum_j |\mathbf{v}_j|^2 f_{eq,j}(\boldsymbol{\alpha}) w_j - 2E_n
\end{bmatrix},
\qquad
\boldsymbol{\alpha} =
\begin{bmatrix}
\alpha_0 \\ \alpha_{1x} \\ \alpha_{1y} \\ \alpha_{1z} \\ \alpha_2
\end{bmatrix},
\end{equation*}
the coefficients are obtained by solving $\mathbf{G}(\boldsymbol{\alpha})=\mathbf{0}$ independently at each configuration-space degree of freedom. GUERNICA solves this nonlinear system with Newton's method, starting from the solution of the previous evaluation at that node. The Jacobian required for the Newton iteration takes the form,
\begin{equation*}
\label{eq:bgk_jac}
J_{mn}(\boldsymbol{\alpha}) =\frac{\partial G_m}{\partial \alpha_n} =\sum_{j=1}^{N_v} \psi_m(\mathbf{v}_j)\, \psi_n(\mathbf{v}_j)\, f_{eq,j}(\boldsymbol{\alpha})\, w_j,
\end{equation*}
where the basis functions $\{\psi_m\}_{m=1}^5$ correspond to the conserved moment functions,
\begin{equation*}
\psi(\mathbf{v}) = \big(1,\ v_x,\ v_y,\ v_z,\ |\mathbf{v}|^2\big).
\end{equation*}
Once $\boldsymbol{\alpha}$ is obtained, the discrete equilibrium $f_{eq,j}$ is evaluated via Eq.~\ref{eq:bgk_exp} and the BGK collision operator is applied pointwise in velocity space as,
\begin{equation*}
C_{bgk}[f_n](\mathbf{v}_j) = -\nu_{nn}\big(f_{n,j}-f_{eq,j}\big).
\end{equation*}
By construction, the discrete operator then conserves mass, momentum and energy under the velocity quadrature, to the tolerance of the Newton solve.

\noindent \textbf{Recombination}

The recombination source defined in Eq.~\ref{eq:Recomb} is evaluated pointwise as,
\begin{equation*}
    S_{rec}(\mathbf{v}_j) = n_e \langle \sigma_{rec} v_e \rangle(T_e) f_{i,j},
\end{equation*}
where the Maxwellian-averaged rate coefficient $\langle \sigma_{rec} v_e \rangle(T_e)$ is evaluated using the radiative recombination fits of Janev et al.~\cite{janev1987}, which depend on $T_e$ only. Since $f_{i,j}$ is normalized so that its quadrature sum equals $n_i$, the velocity integral of the discrete source equals the recombination rate $n_e n_i \langle \sigma_{rec} v_e \rangle$ exactly.

\subsection{Time advancement and steady-state solution}
\label{subsec:Time}

The DG--DVM discretization yields the semi-discrete system
\begin{equation}
\label{eq:semiDiscrete}
    \frac{d \mathbf{F}}{dt} = \mathcal{L}(\mathbf{F}),
\end{equation}
where $\mathbf{F}$ contains all configuration-space degrees of freedom for all discrete velocities, and $\mathcal{L}$ denotes the combined contribution of DG transport, collision operators, boundary fluxes and explicit source terms. GUERNICA supports two modes of advancement for this system: transient and steady-state mode. All terms are advanced explicitly in both modes.

In transient mode, GUERNICA advances Eq.~\ref{eq:semiDiscrete} with a third-order strong stability preserving Runge--Kutta (SSP-RK3) method. Since the DG transport operator is treated explicitly, the time step is constrained by a Courant--Friedrichs--Lewy (CFL) condition associated with the DG discretization of the advection operator. We measure the advective rate of velocity $\mathbf{v}_j$ in element $K$ by the outgoing flux per unit area,
\begin{equation}
\label{eq:gamma}
\gamma^K_j = \frac{1}{|K|}\sum_{e\in\partial K} |e|\,\max\left(0,\mathbf{v}_j\cdot\mathbf{n}_e\right),
\end{equation}
where the sum runs over the four edges $e$ of $K$, $|e|$ is the edge length, $\mathbf{n}_e$ is the outward unit normal and $|K|$ is the element area. For a rectangular element this reduces to $|v_{j,x}|/h^K_{x} + |v_{j,y}|/h^K_{y}$, with $h^K_{x}$ and $h^K_{y}$ the element sizes. A useful estimate of the stable time step is then
\begin{equation}
\label{eq:CFL}
\Delta t \lesssim \frac{1}{2p+1}\left(\max_{K,j}\gamma^K_j\right)^{-1},
\end{equation}
where $p$ is the DG polynomial degree and the maximum is taken over elements and discrete velocities \cite{cockburn1998}. The time step is a fixed input in transient mode. The explicit collision terms impose a further limit, since $\nu\Delta t$ must lie within the stability region of the integrator for every local loss frequency $\nu$.

For neutral transport on a static background, the quantity of interest is generally the converged steady-state distribution rather than the physical transient. In steady-state mode, GUERNICA therefore solves the steady-state equation
\begin{equation}
\label{eq:steady_residual}
\mathcal{L}(\mathbf{F}) = 0
\end{equation}
by pseudo-time iteration. Each iteration is a single forward-Euler step with a local pseudo-time step,
\begin{equation*}
\label{eq:steady_local_update}
\mathbf{F}^{K,m+1}_j = \mathbf{F}^{K,m}_j + \Delta\tau^K_j\,\mathcal{L}^K_j(\mathbf{F}^{m}),
\end{equation*}
where $m$ denotes the pseudo-time iteration index, $\mathbf{F}^{K}_j$ and $\mathcal{L}^K_j$ are the blocks of the state and of the right-hand side belonging to element $K$ and velocity $\mathbf{v}_j$, and $\Delta\tau^K_j$ is a pseudo-time step assigned independently to each element--velocity pair.

The local step is set from the advective rate of Eq.~\ref{eq:gamma},
\begin{equation*}
\label{eq:local_dtau}
\Delta \tau^K_{j} = \frac{\Delta\tau_0}{\max\left(1,\gamma^K_j\right)},
\end{equation*}
where $\Delta\tau_0$ is a single input parameter. Wherever $\gamma^K_j \geq 1$ in dimensionless units, $\Delta\tau_0$ is the local Courant number $\gamma^K_j\Delta\tau^K_j$. Elsewhere, $\Delta\tau_0$ is the step itself, which keeps the step bounded for the slowest velocities. This allows each element--velocity block to be relaxed according to its local advective time scale. Only the in-plane velocity components enter $\gamma^K_j$, so the step is the same for all out-of-plane velocities. The explicit collision terms impose a second limit, $\nu\Delta\tau^K_j < 2$ for a local loss frequency $\nu$, which Sec.~\ref{sec:stability} quantifies.

Local stepping is not without cost. The BGK operator is constructed so that its discrete velocity moments vanish, but applying a different pseudo-time step to each velocity destroys this cancellation. A moment-preserving pseudo-time treatment is under development and is left to future work as discussed in Sec.~\ref{sec:Conc}. The results of this paper therefore include neutral--neutral collisions only in transient mode with the exception of Sec.~\ref{sec:cost} to measure its cost consistent with other operators.

We monitor convergence with three scalar residuals formed from $\mathcal{L}(\mathbf{F}^m)$ before it is scaled by the local step. A global relative norm of the distribution function monitors convergence of the velocity-space structure,
\begin{equation*}
\label{eq:f_raw_res}
r_f^m = \frac{\|\mathcal{L}(\mathbf{F}^m)\|_2}{\|\mathbf{F}^m\|_2},
\end{equation*}
while the remaining two residuals are taken on the density and energy moments of $\mathcal{L}(\mathbf{F}^m)$, normalized element by element and combined by a maximum over the mesh,
\begin{equation*}
\label{eq:moment_raw_res}
r_n^m = \max_K \frac{\left\|\sum_j w_j \mathcal{L}^K_j(\mathbf{F}^m)\right\|_K}{\left\|\sum_j w_j \mathbf{F}^{K,m}_j\right\|_K},
\qquad
r_E^m = \max_K \frac{\left\|\sum_j w_j |\mathbf{v}_j|^2 \mathcal{L}^K_j(\mathbf{F}^m)\right\|_K}{\left\|\sum_j w_j |\mathbf{v}_j|^2 \mathbf{F}^{K,m}_j\right\|_K}.
\end{equation*}
Here $\|\cdot\|_2$ is the Euclidean norm over all phase-space degrees of freedom and $\|\cdot\|_K$ is the Euclidean norm over the $N_p$ nodal values of element $K$, so each moment residual is normalized by the same moment of the state. This ensures that low-density regions are not left unconverged simply because the criterion is dominated by the region of largest density as would be the case if only the global norm were considered. The iteration stops when all three residuals fall below a prescribed tolerance or the total number of iterations hits a prescribed max.

\section{Verification and validation}
\label{sec:VV}

This section verifies the implementation of GUERNICA and characterizes its convergence, computational scaling and accuracy. We first verify the individual operators in isolation, followed by transport and the wall boundary condition against exact solutions. After examining the stability of the steady-state solver and the cost of each operator, we validate ionization and charge exchange together against the Monte Carlo code DEGAS2 in the one-dimensional benchmark of Bernard et al.~\cite{bernard2022}. The recombination source has no standalone test; its discrete normalization and injection share the source kernel verified in Sec.~\ref{sec:transport}, but its rate coefficient is not independently verified here. Unless stated otherwise, errors are relative $L^2$ norms of the neutral density, and the one-dimensional problems are solved on two-dimensional meshes that are uniform and periodic in the transverse direction. Quantities given without units are in the dimensionless units of Sec.~\ref{subsec:Dims}, with $T_0 = 1$~eV, $L_0 = 1$~m and $n_0 = 10^{18}$~m$^{-3}$.

\subsection{Operator verification}
\label{sec:operators}

\begin{figure}[!htbp]
\centering
\begin{subfigure}[t]{0.48\textwidth}\centering
\includegraphics[width=\textwidth]{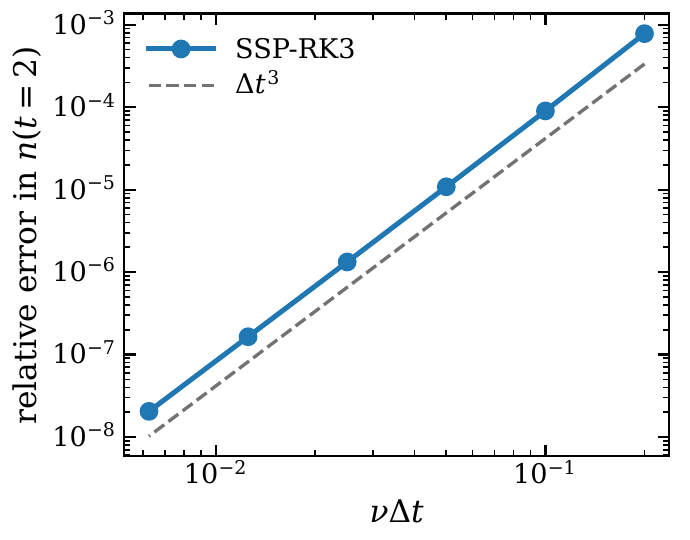}
\caption{Ionization decay}
\end{subfigure}\hfill
\begin{subfigure}[t]{0.48\textwidth}\centering
\includegraphics[width=\textwidth]{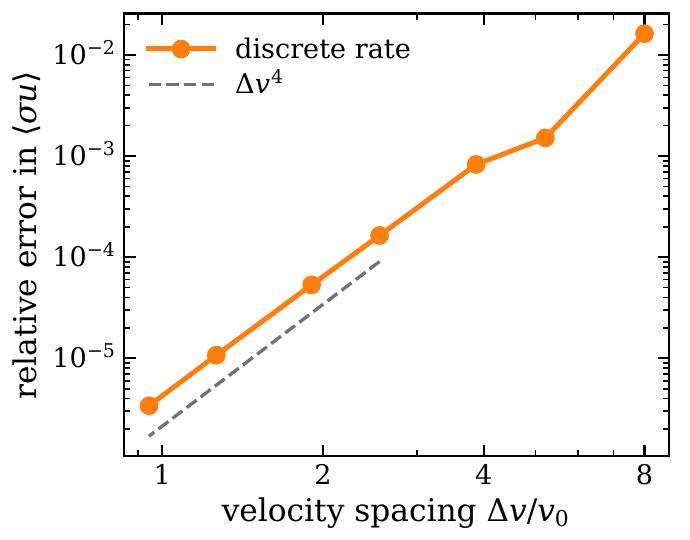}
\caption{Charge exchange rate}
\end{subfigure}
\caption{Spatially homogeneous verification. (a) Error in the density of a uniformly ionizing Maxwellian at $t = 2$ against the time step. (b) Error in the charge exchange rate between 10~eV neutrals and 60~eV drifting ions as the velocity grid is refined in all three directions.}
\label{fig:homogeneous}
\end{figure}

We verify the ionization operator and the SSP-RK3 time integrator together using a spatially uniform Maxwellian subject to a constant ionization rate $\nu = 1$. Streaming does not act on a uniform state, so the density decays exactly as $n(t) = n(0)e^{-\nu t}$ and any error comes from the time integration and ionization implementation. We advanced the problem to $t = 2$, and as seen in Fig.~\ref{fig:homogeneous}a, the error decreases at third order, with an observed rate of 3.01 at the smallest step. For this linear problem each SSP-RK3 step multiplies the density by $1 - z + z^2/2 - z^3/6$ with $z = \nu\Delta t$, and the computed density matches this polynomial to round-off ($<3\times10^{-14}$) at every step size, confirming that both the ionization operator and the integrator are implemented as intended.

For charge exchange there is no simple exact solution of the full problem, but the reaction rate between two Maxwellian populations is known. We consider neutrals at 10~eV and ions at 60~eV drifting at $5v_0$ and compare the discrete loss rate with $n_n n_i\langle\sigma u\rangle$, where $u$ is the relative speed and the average over the Maxwellian distribution of relative velocities is evaluated by high-order quadrature to a relative accuracy of about $10^{-13}$. We refine all three velocity directions together over $[-60, 60]^3~v_0$. The error, shown in Fig.~\ref{fig:homogeneous}b, falls from $1.6\times10^{-2}$ with 16 points per direction to $3.4\times10^{-6}$ with 128, and the observed rate approaches four (3.96 at the finest level). Fourth order is the rate expected for a lattice rule in three dimensions when the integrand has a conical singularity at a grid point \cite{lyness1976}, as the kernel $K_{cx}$ does at zero relative velocity. Further unit tests of the operator confirm that it conserves particles at every spatial point, that it vanishes when $f_n \propto f_i$ and that the FFT evaluation reproduces the direct evaluation of the convolution, all to machine precision in double precision.

\begin{figure}[!htbp]
\centering
\begin{subfigure}[t]{0.48\textwidth}\centering
\includegraphics[width=\textwidth]{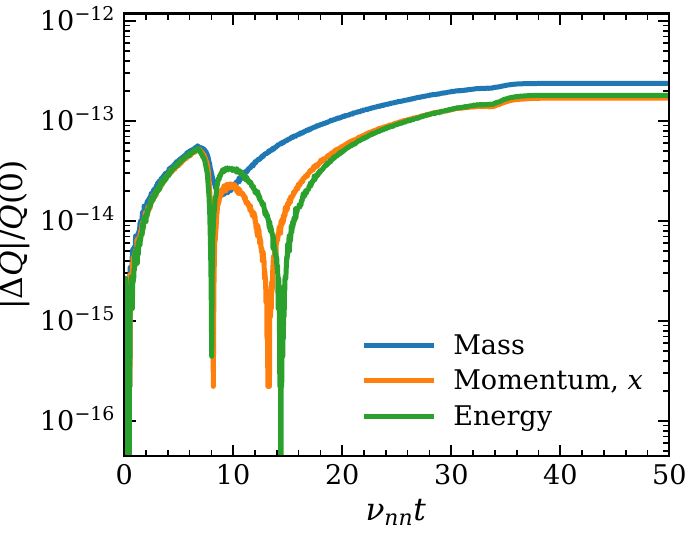}
\caption{Mass, momentum and energy conservation}
\end{subfigure}
\begin{subfigure}[t]{0.48\textwidth}\centering
\includegraphics[width=\textwidth]{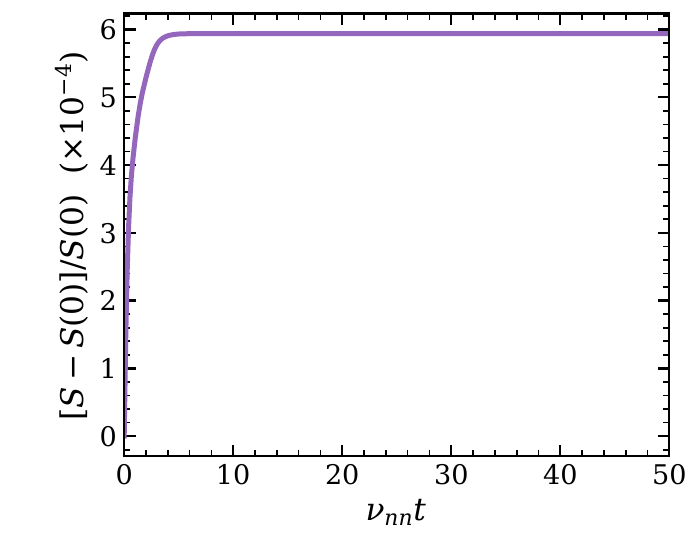}
\caption{Entropy}
\end{subfigure}
\caption{Verification of the BGK operator. (a) Relative change of mass, momentum in $x$ and energy in time, showing that all quantities are conserved up to round-off error. (b) Relative change of entropy in time, showing that it increases monotonically and asymptotes as it reaches equilibrium.}
\label{fig:bgk}
\end{figure}

We verify the BGK operator on a periodic domain $x\in[0, 1]$, evolving the kinetic equation with only spatial transport and BGK collisions at a constant frequency $\nu_{nn} = 5$. The initial condition is a uniform Maxwellian perturbed by an anisotropic component,
\begin{equation*}
f(x, \mathbf{v}, 0) = f_M(\mathbf{v})\left[1 + \epsilon\cos(2\pi x)\,\xi(\mathbf{v})\right],
\end{equation*}
where $f_M$ has density $n_0 = 1$, bulk velocity $\mathbf{u}_0 = (1, 0, 0)$ and temperature $T_0 = 1$, $\epsilon = 0.1$ and 
\begin{equation*}
    \xi(\mathbf{v}) = \frac{(v_x - u_{0_x})^2 - (v_y - u_{0_y})^2}{2T_0}.
\end{equation*}
Under the Maxwellian weight, $\xi$ is orthogonal to $1$, $\mathbf{v}$ and $|\mathbf{v}|^2$, so the perturbation carries no mass, momentum or energy, and BGK relaxation damps it without changing the conserved moments. We discretized the domain with four elements of degree $p = 4$ and advanced the solution with SSP-RK3 at $\Delta t = 10^{-3}$ to $\nu_{nn}t = 50$, with the velocity domain $[-10, 10]^3$ and $N_{vx} = N_{vy} = N_{vz} = 77$. As seen in Fig.~\ref{fig:bgk}a, the relative changes in mass, $x$-momentum and energy remain below $2.4\times10^{-13}$, growing only through round-off accumulation, with the dips corresponding to sign changes of the deviation. The discrete entropy, $S = -\int_{\Omega_x} \sum_j w_j f_{n,j}\ln f_{n,j}\, d\mathbf{x}$, shown in Fig.~\ref{fig:bgk}b, increases monotonically and levels off as the perturbation is damped. These results verify that the BGK operator conserves mass, momentum and energy to round-off under the discrete velocity quadrature and drives the system monotonically toward equilibrium.

\subsection{Transport tests}
\label{sec:transport}

We verify transport and the boundary conditions using problems in which, with charge exchange turned off and a constant ionization rate $\nu$, every discrete velocity decouples. The exact solution can then be written both for the continuous problem and for the discrete-velocity problem that GUERNICA actually solves. The difference between GUERNICA and the discrete-velocity solution measures the error of the DG discretization, and the difference between the discrete and continuous solutions measures the error of the velocity discretization.

\begin{figure}[!htbp]
\centering
\begin{subfigure}[t]{0.48\textwidth}\centering
\includegraphics[width=\textwidth]{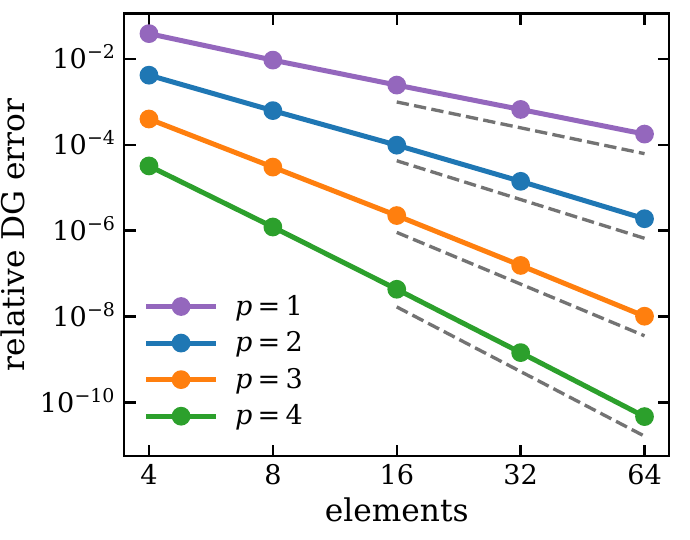}
\caption{Spatial convergence}
\end{subfigure}\hfill
\begin{subfigure}[t]{0.48\textwidth}\centering
\includegraphics[width=\textwidth]{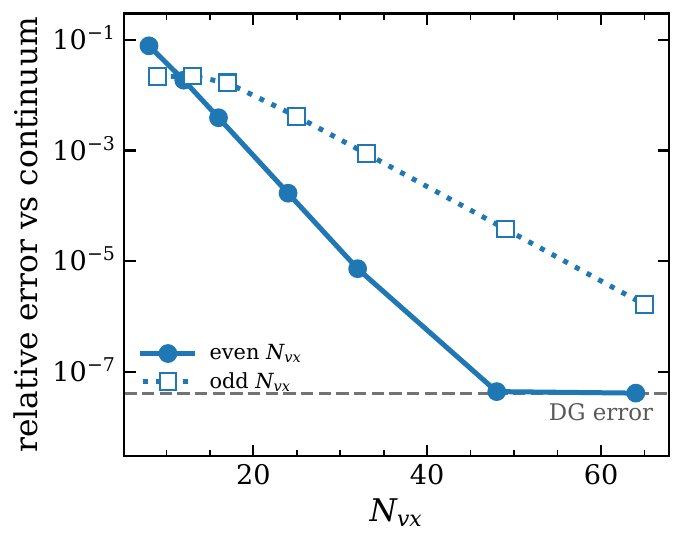}
\caption{Velocity quadrature}
\end{subfigure}
\caption{Periodic source test. (a) DG error against the number of elements for $p = 1$ to 4. (b) Error against the continuous solution as the number of velocity points in $v_x$ is increased, for even and odd numbers of points, at $p = 4$ on 16 elements and $\nu/k = 1$. The dashed line is the DG error of this mesh.}
\label{fig:b1}
\end{figure}

The first test is a periodic domain of length 1 with a source $S = [c_0 + c_1\cos kx]\,M(\mathbf{v})$, where $M$ is a Maxwellian with $T = 1$, $k = \nu = 2\pi$, $c_0 = 1$ and $c_1 = 0.5$. The steady state density is $c_0/\nu + c_1 A\cos kx$, where the amplitude of the continuous problem is
\begin{equation*}
A = \sqrt{\frac{\pi}{2}}\,\frac{1}{k\sqrt{T}}\operatorname{erfcx}\!\left(\frac{\nu}{k\sqrt{2T}}\right),
\end{equation*}
and that of the discrete-velocity problem is the corresponding sum over the velocity points. A detailed derivation is given in~\ref{app:fourier}. We used uniform meshes of 4 to 64 elements for $p = 1$ to 4, with 32 velocity points in $v_x$ over $[-8, 8]$ and a single transverse velocity. Figure~\ref{fig:b1}a shows that the DG error converges at the expected $O(\Delta x^{p+1})$ rate, with observed rates of 1.91, 2.89, 3.93 and 4.95 for $p = 1$ to 4 on the finest meshes, and errors down to $4.7\times10^{-11}$. We repeated the test in two dimensions with the wave oblique to the mesh, on meshes whose interior vertices were displaced randomly by up to 25\% of the element size so that no element is a parallelogram. The observed rates were approximately the same as on regular meshes (not shown), showing that the formulation retains its accuracy on general bilinear quadrilaterals such as those of the MAST-U mesh in Sec.~\ref{sec:MAST}.

The same problem also illustrates the velocity quadrature of Sec.~\ref{subsec:DVM}. For an even number of points in a direction the quadrature weights are uniform away from the ends of the grid, and for smooth distributions that have decayed at the edges of the velocity domain the error decreases exponentially with the number of points. For an odd number of points the composite Simpson weights alternate, and the error also decreases exponentially but at half the rate. As seen in Fig.~\ref{fig:b1}b, at $p = 4$ on 16 elements, the total error with even $N_{vx}$ falls from $3.9\times10^{-3}$ at 16 points to the level of the DG error at 48, after which it stays at that level because the mesh is fixed, while with odd $N_{vx}$ it is still $1.7\times10^{-6}$ at 65. For this reason even numbers of velocity points are used in the remainder of the paper wherever the distribution must be resolved.

\begin{figure}[!htbp]
\centering
\begin{subfigure}[t]{0.48\textwidth}\centering
\includegraphics[width=\textwidth]{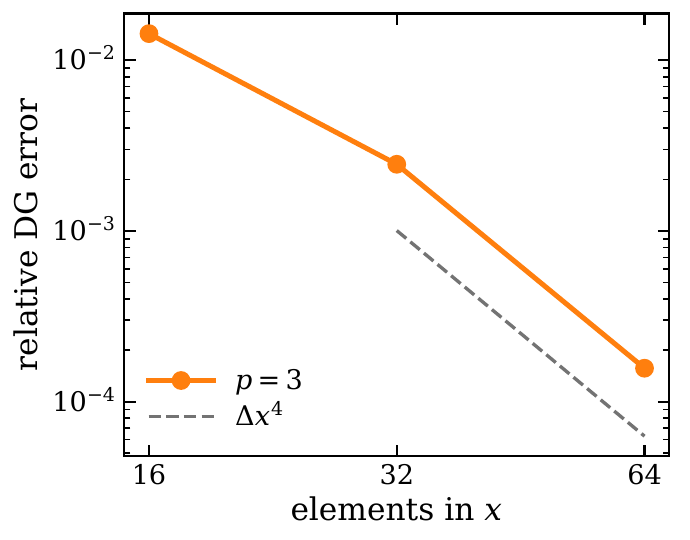}
\caption{Spatial convergence}
\end{subfigure}\hfill
\begin{subfigure}[t]{0.48\textwidth}\centering
\includegraphics[width=\textwidth]{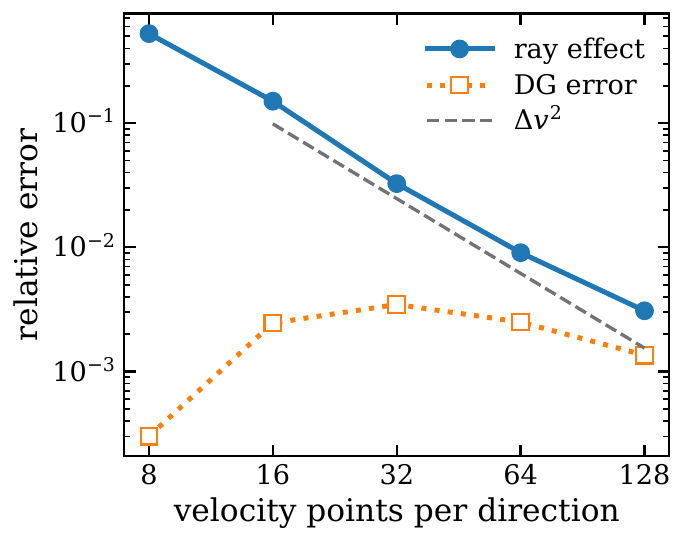}
\caption{Ray effect}
\end{subfigure}
\caption{Emitting wall segment. (a) DG error against the number of elements along $x$, with $16\times16$ velocity points and $p = 3$. (b) Ray effect and DG error against the number of velocity points per direction, on $32\times16$ elements.}
\label{fig:b3}
\end{figure}

We verify the diffuse wall boundary condition of Sec.~\ref{subsec:BCNum} on a $4\times2$ domain whose walls absorb all incoming neutrals, except for a segment of width 2 on the lower wall that emits a Maxwellian with $T_w = 1$ and a spatially varying amplitude $\alpha(x) = \cos^6(\pi(x - 2)/2)$, into a medium with $\nu = 0.5$. The problem has two velocity dimensions, with $v_x$ and $v_y$ spanning $[-6, 6]$ at every resolution. Each discrete velocity can be traced back along its characteristic to the wall, which gives the discrete-velocity solution, while the continuous solution is an integral over the emission angle. As seen in Fig.~\ref{fig:b3}a, with $16\times16$ velocity points and $p = 3$, the DG error falls from $1.4\times10^{-2}$ to $1.6\times10^{-4}$ as the mesh is refined from 16 to 64 elements along the wall, with an observed rate of 3.97 on the finest pair, verifying the implementation of the wall condition.

The difference between the discrete-velocity and continuous solutions in this problem is known as the ray effect. A finite set of velocity directions samples a source of limited angular extent, so the density far from the source shows streaks along the directions of the velocity grid \cite{lathrop1968}. With $32\times16$ elements, the relative $L^2$ difference between the two solutions over the domain is 53\%, 15\%, 3.3\%, 0.9\% and 0.3\% with $8^2$, $16^2$, $32^2$, $64^2$ and $128^2$ velocity points, decreasing at roughly second order in the velocity spacing, while the DG error remains below $3.5\times10^{-3}$ at every velocity resolution and so stays small compared with the ray effect until both fall below 1\% as seen in Fig.~\ref{fig:b3}b. At the lower resolutions the ray effect is by far the largest error in the solution.

\subsection{Stability and convergence of the steady-state solver}
\label{sec:stability}

A higher polynomial degree gives a higher order of convergence but also restricts the pseudo-time step of the steady-state solver of Sec.~\ref{subsec:Time}. To quantify this, we solved the high-density benchmark of Sec.~\ref{sec:degas2} with $(62, 18, 18)$ velocity points while scanning the plasma density $n_e$ from $5\times10^{16}$ to $5\times10^{19}$~m$^{-3}$. Every reaction rate is proportional to $n_e$, so the only parameter that changes is the collisionality, measured as the ratio of the domain length to the neutral mean free path, $L/\lambda \approx 20\,n_e/(10^{18}\,\mathrm{m}^{-3})$, where $\lambda$ is the thermal speed of the 10~eV source neutrals divided by the sum of the ionization frequency and the Maxwellian-averaged charge exchange frequency. At each density and for $p = 1$ to 4, we ran a sequence of values of the pseudo-time parameter $\Delta\tau_0$ to determine the stability boundary.

\begin{figure}[!htbp]
\centering
\begin{subfigure}[t]{0.48\textwidth}\centering
\includegraphics[width=\textwidth]{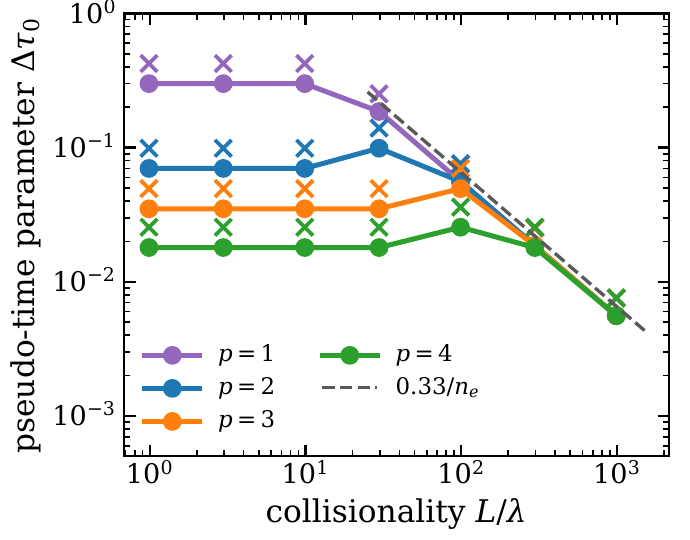}
\caption{Stability limit}
\end{subfigure}\hfill
\begin{subfigure}[t]{0.48\textwidth}\centering
\includegraphics[width=\textwidth]{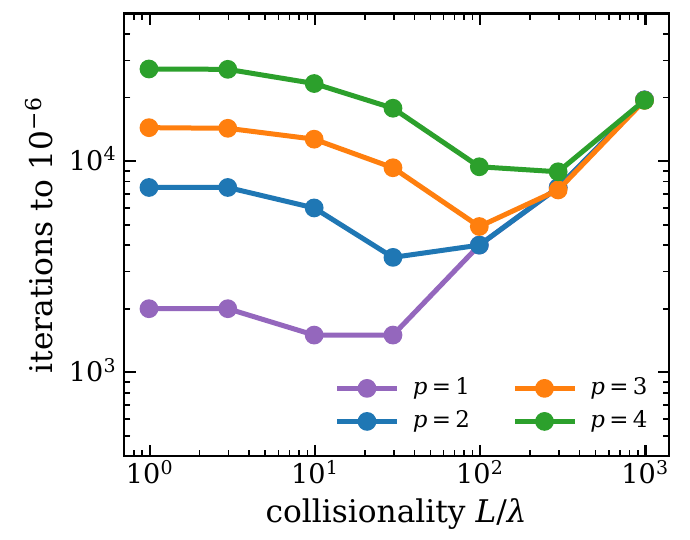}
\caption{Iterations to convergence}
\end{subfigure}
\caption{Steady-state solver against collisionality for $p = 1$ to 4. (a) Largest stable (filled) and smallest unstable ($\times$) values of the pseudo-time parameter $\Delta\tau_0$, with the collisional limit $0.33/n_e$. (b) Iterations to bring all three residuals below $10^{-6}$ at about half of the largest stable value.}
\label{fig:stability}
\end{figure}

Figure~\ref{fig:stability}a shows the largest stable value, where two limits appear. At low collisionality the limit does not depend on density and is set by transport: because $\gamma^K_j \geq 1$ everywhere on this grid, $\Delta\tau_0$ is the local Courant number of every element--velocity pair, as discussed in Sec.~\ref{subsec:Time}. This limit falls rapidly with the polynomial degree, from about 0.36 at $p = 1$ to 0.08, 0.035 and 0.018 at $p = 2$, 3 and 4, roughly halving with each order above $p = 2$. At high collisionality the limit is set by the collision terms, since the forward-Euler update of a local loss at rate $\nu$ is stable only for $\nu\Delta\tau < 2$. The local step depends only on the in-plane velocity, while the out-of-plane velocity raises the charge exchange rate. The most restrictive cells are therefore those of the slowest in-plane velocities with the largest $|v_z|$ in the widest element, where $\Delta\tau = \Delta\tau_0/1.52$. For this problem this gives $\Delta\tau_0 < 0.33/n_e$, which agrees with the measured limit at every polynomial degree.

Figure~\ref{fig:stability}b shows the number of iterations needed to bring all three residuals below $10^{-6}$ at about half of the largest stable value of $\Delta\tau_0$. At high collisionality, where the collisional limit applies, the number of iterations is the same for every polynomial degree and grows with the density, reaching about 19,500 at $L/\lambda = 1000$. At low collisionality the smaller step at higher order translates directly into more iterations, from about 2,000 at $p = 1$ to 7,500, 14,000 and 27,000 at $p = 2$, 3 and 4. Since the cost of an iteration per spatial node is nearly independent of $p$, as discussed in Sec.~\ref{sec:cost}, the choice of polynomial degree is a trade between accuracy per node and the number of iterations, and this trade is most pronounced in weakly collisional regions.

\subsection{Computational cost}
\label{sec:cost}

\begin{figure}[!htbp]
\centering
\begin{subfigure}[t]{0.48\textwidth}\centering
\includegraphics[width=\textwidth]{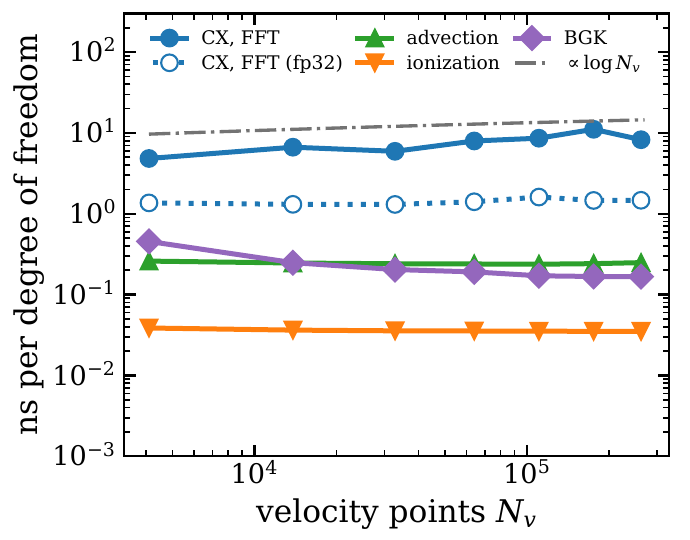}
\caption{Velocity resolution}
\end{subfigure}\hfill
\begin{subfigure}[t]{0.48\textwidth}\centering
\includegraphics[width=\textwidth]{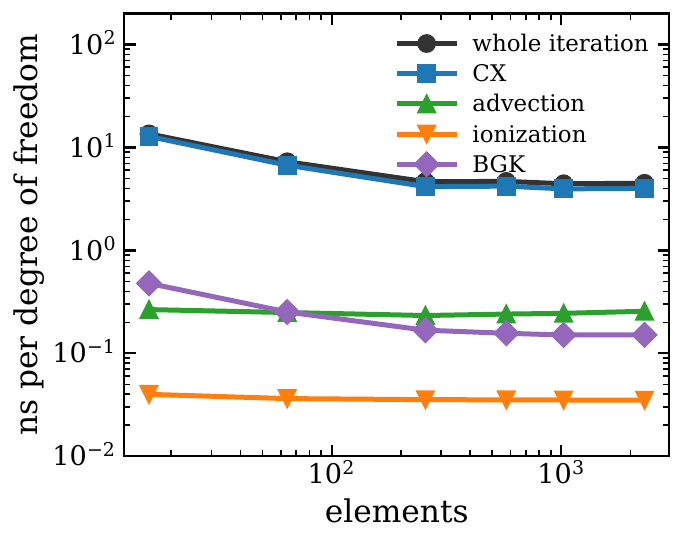}
\caption{Number of elements}
\end{subfigure}
\caption{Operator cost on one RTX A5000, per phase-space degree of freedom. (a) Time per evaluation against the number of velocity points, $8\times8$ elements and $p = 2$, for charge exchange with the FFT in double and single precision, advection, ionization and the BGK operator. The reference line is proportional to $\log N_v$, corresponding to an $O(N_v\log N_v)$ cost. (b) Time per iteration against the number of elements, $p = 2$ and $24^3$ velocity points.}
\label{fig:timing}
\end{figure}

We measured the cost of each operator on a single NVIDIA RTX A5000 using device-synchronized timers. We ran each case for 200 steady-state iterations using the plasma of the low-density benchmark of Sec.~\ref{sec:degas2} ($n = 5\times10^{18}$~m$^{-3}$, no ion flow) in a 1~m square box with periodic boundaries. The base case uses $8\times8$ elements with $p = 2$ and $24^3$ velocity points, and the velocity resolution, the number of elements and the polynomial degree were varied independently from it. The results are shown in Fig.~\ref{fig:timing}.

Charge exchange dominates the cost, taking 90--95\% of each iteration in double precision. With the FFT evaluation, its cost grows from 11.5~ms at $16^3$ velocity points to 1.25~s at $64^3$, following $N_v^{1.13}$, close to the expected $O(N_v\log N_v)$.  The direct evaluation of the convolution grows as $N_v^{2.02}$, consistent with $O(N_v^2)$. It is already 83 times slower than the double precision FFT at $16^3$ velocity points, and at $48^3$ a single evaluation takes 12 minutes against 0.55~s with the FFT, a factor of 1,300, so the resolutions used in this paper would not be practical without the FFT.\footnote{The direct evaluation is omitted from Fig.~\ref{fig:timing}a, where it would lie two to three orders of magnitude above every other curve.} Evaluating the convolution in single precision, as described in Sec.~\ref{subsec:ColNum}, makes the operator 3.5--7.6 times faster with minimal impact on the solution, as examined in Secs.~\ref{sec:degas2} and \ref{sec:MAST}.

For the BGK operator, the sums in the Newton Jacobian of Sec.~\ref{subsec:ColNum} factorize on the tensor-product velocity grid, so the Newton solve costs $O(N_{vx} + N_{vy} + N_{vz})$ per spatial node, and it converged in at most three iterations at every resolution tested. At $64^3$ the complete BGK operator costs about 2\% of the charge exchange operator. Advection, ionization and the volume source, which shares its kernel with recombination, perform a fixed amount of work per spatial degree of freedom and velocity, and scale linearly with $N_v$. 

Above roughly 256 elements for this problem, the GPU is saturated and the cost is linear in the number of elements. The cost per phase-space degree of freedom then levels off at about 4.5~ns per iteration in double precision, as seen in Fig.~\ref{fig:timing}b. The cost of charge exchange per spatial node varies by less than a factor of two from $p = 1$ to 5, with no systematic trend in $p$ (not shown), and while the cost of advection per node grows with $p$ it remains a small fraction of the total, so the cost per iteration is set mainly by the total number of spatial nodes and velocity points.

\subsection{Comparison with DEGAS2}
\label{sec:degas2}

We validate the combined ionization and charge exchange model against DEGAS2 using the one-dimensional benchmark of Bernard et al.~\cite{bernard2022}. The benchmark has a low-density and a high-density case. In both, the hydrogen plasma has constant temperatures $T_e = 30$~eV and $T_i = 60$~eV on $x\in[0, L_x]$, with a prescribed flow along the streaming direction $x$,
\begin{equation*}
\mathbf{u}_e(x) = \mathbf{u}_i(x) = u_i(x)\,\hat{\mathbf{x}}, \qquad 
u_i(x) = c_{s0}\frac{x - L_x/2}{L_x/2}, \qquad
c_{s0} = \sqrt{\frac{T_e}{m_i}},
\end{equation*}
where $c_{s0}$ is the reference sound speed. The two cases differ in density and domain length: the low-density case has constant densities $n_e = n_i = 5\times10^{18}$~m$^{-3}$ over $L_x = 40$~m, and the high-density case $n_e = n_i = 5\times10^{19}$~m$^{-3}$ over $L_x = 10$~m. In both cases the initial neutral condition, the volumetric neutral source, plasma profiles and boundary conditions follow the benchmark definition~\cite{bernard2022}. Ionization uses the effective rates of the benchmark, $\langle\sigma_{iz}v_e\rangle = 2.147\times10^{-14}$ m$^3$/s in the low-density case and $2.595\times10^{-14}$~m$^3$/s in the high-density case, and charge exchange uses the same combination of the Krsti\'c \& Schultz and Janev \& Smith fits as DEGAS2.

GUERNICA discretizes configuration space with graded $p = 4$ elements, 72 in the low-density case and 76 in the high-density case. The smallest elements (5~mm in the high-density case) are at the walls. Velocity space uses $(N_{vx}, N_{vy}, N_{vz}) = (82, 24, 24)$ points spanning $\pm3.0\times10^5$~m/s, where $v_x$ is the streaming direction. DEGAS2 solves the problem in full three-dimensional velocity space with 896 (low density) and 1792 (high density) zones, within which the plasma is constant and the neutral quantities are cell-averaged Monte Carlo tallies. The differences quoted below therefore use the GUERNICA density and density-weighted temperature averaged over each DEGAS2 zone, while the figures show the GUERNICA solution at its nodes. On one RTX A5000 GPU with the double-precision convolution, GUERNICA brings all three residuals below $10^{-7}$ in 1.5~h (7,200 iterations) for the low-density case and below $10^{-9}$ in 6~h (27,200 iterations) for the high-density case.

\begin{figure}[!htbp]
\centering
\begin{subfigure}[t]{0.48\textwidth}\centering
\includegraphics[width=\textwidth]{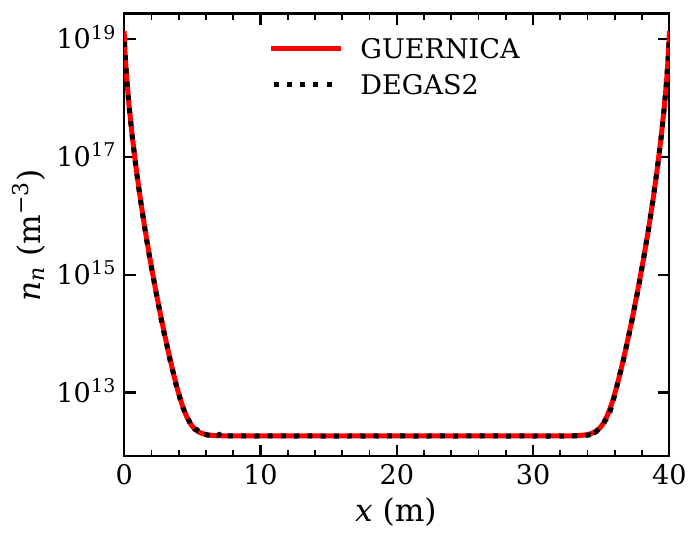}
\caption{Low density, neutral density}
\end{subfigure}\hfill
\begin{subfigure}[t]{0.48\textwidth}\centering
\includegraphics[width=\textwidth]{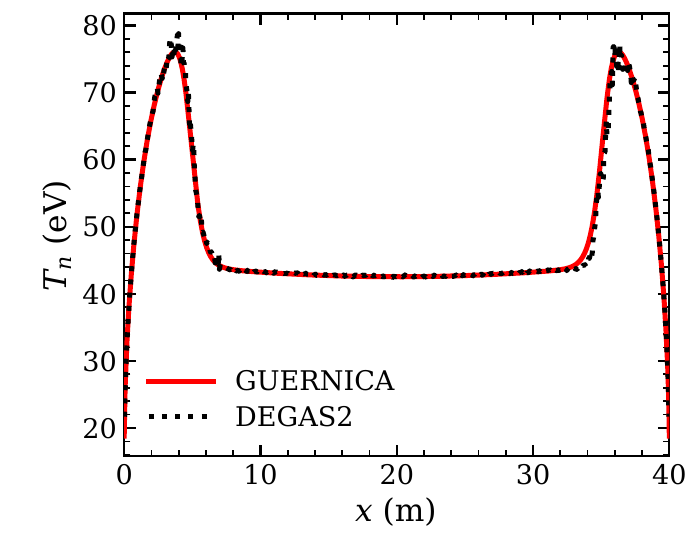}
\caption{Low density, neutral temperature}
\end{subfigure}\\[4pt]
\begin{subfigure}[t]{0.48\textwidth}\centering
\includegraphics[width=\textwidth]{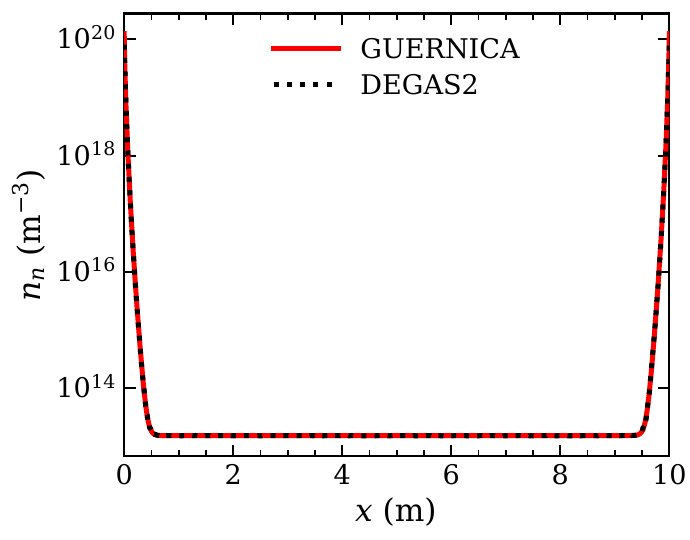}
\caption{High density, neutral density}
\end{subfigure}\hfill
\begin{subfigure}[t]{0.48\textwidth}\centering
\includegraphics[width=\textwidth]{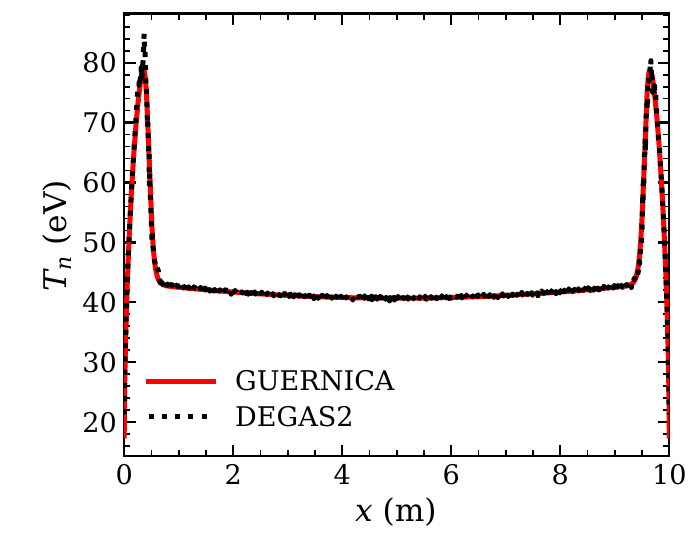}
\caption{High density, neutral temperature}
\end{subfigure}
\caption{Neutral density and temperature from GUERNICA and DEGAS2 for the (a, b) low-density and (c, d) high-density benchmarks with ionization and charge exchange.}
\label{fig:degas2}
\end{figure}

\begin{figure}[!htbp]
\centering
\begin{subfigure}[t]{0.48\textwidth}\centering
\includegraphics[width=\textwidth]{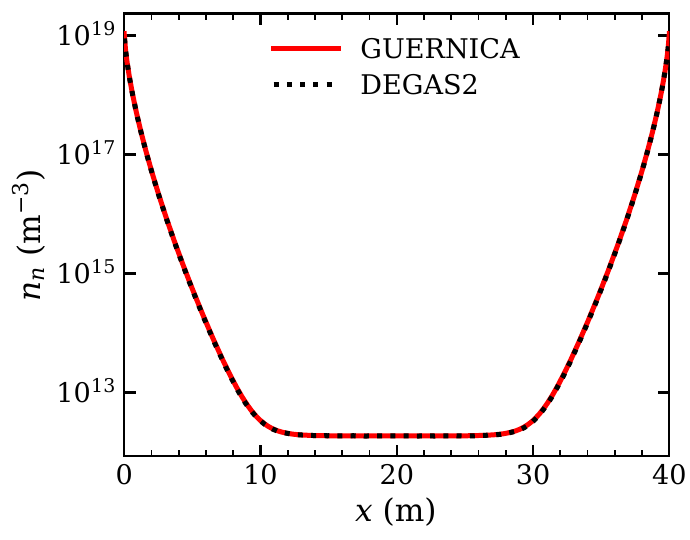}
\caption{Low density, neutral density}
\end{subfigure}\hfill
\begin{subfigure}[t]{0.48\textwidth}\centering
\includegraphics[width=\textwidth]{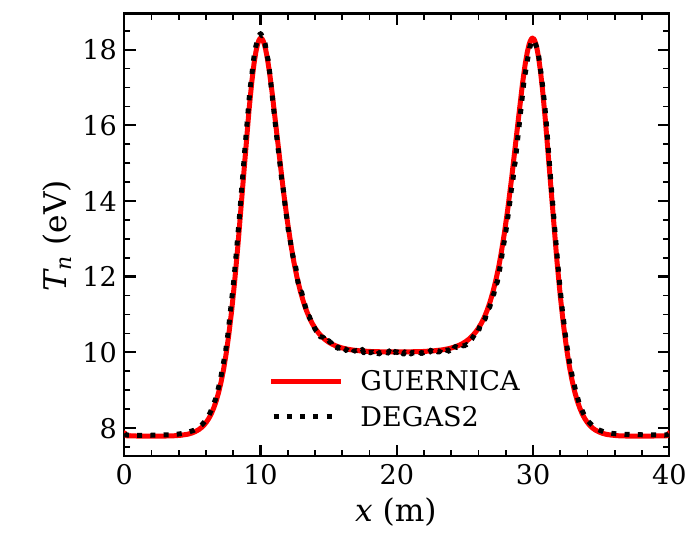}
\caption{Low density, neutral temperature}
\end{subfigure}\\[4pt]
\begin{subfigure}[t]{0.48\textwidth}\centering
\includegraphics[width=\textwidth]{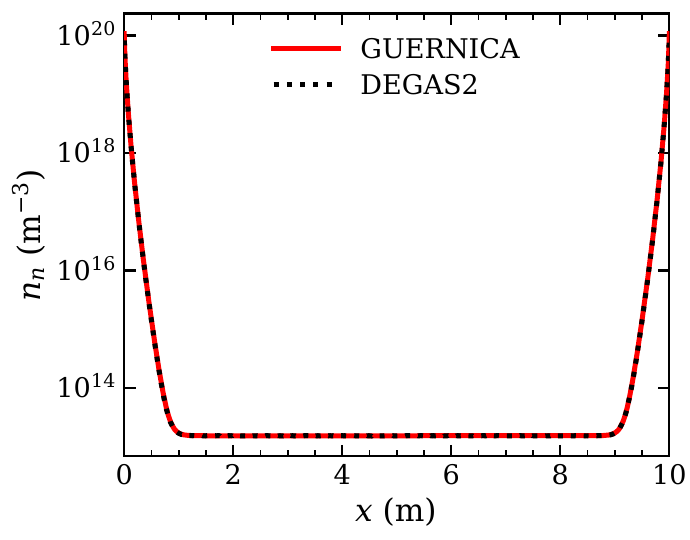}
\caption{High density, neutral density}
\end{subfigure}\hfill
\begin{subfigure}[t]{0.48\textwidth}\centering
\includegraphics[width=\textwidth]{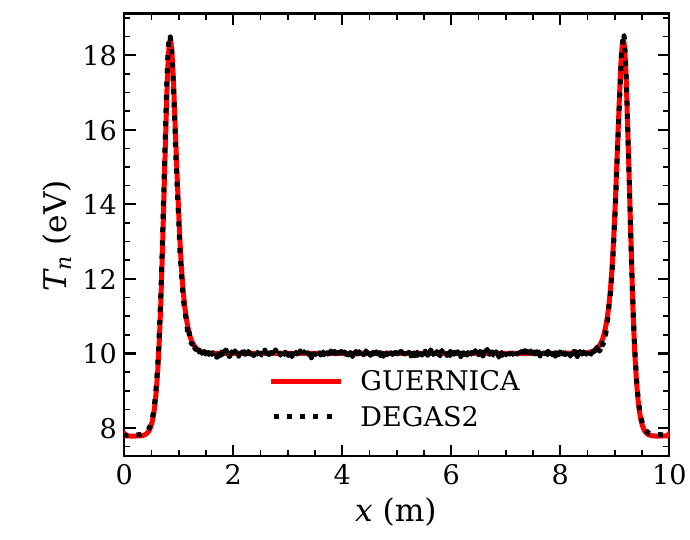}
\caption{High density, neutral temperature}
\end{subfigure}
\caption{Neutral density and temperature from GUERNICA and DEGAS2 for the (a, b) low-density and (c, d) high-density benchmarks with only ionization enabled.}
\label{fig:degas2_iz}
\end{figure}

\begin{figure}[!htbp]
\centering
\begin{subfigure}[t]{0.48\textwidth}\centering
\includegraphics[width=\textwidth]{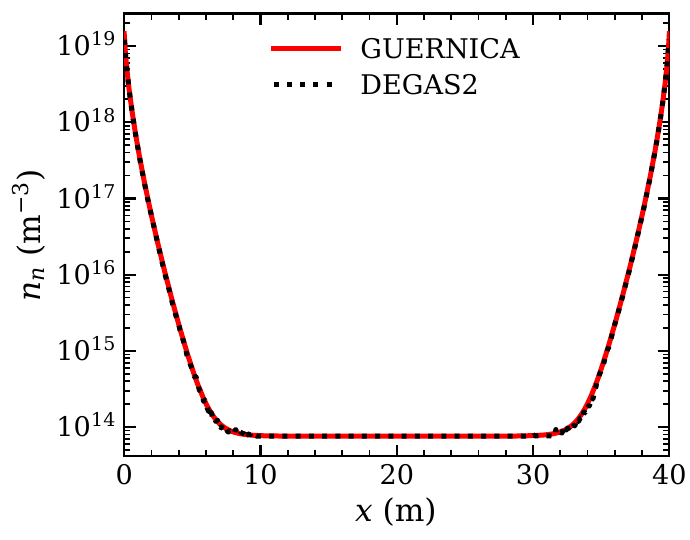}
\caption{Low density, neutral density}
\end{subfigure}\hfill
\begin{subfigure}[t]{0.48\textwidth}\centering
\includegraphics[width=\textwidth]{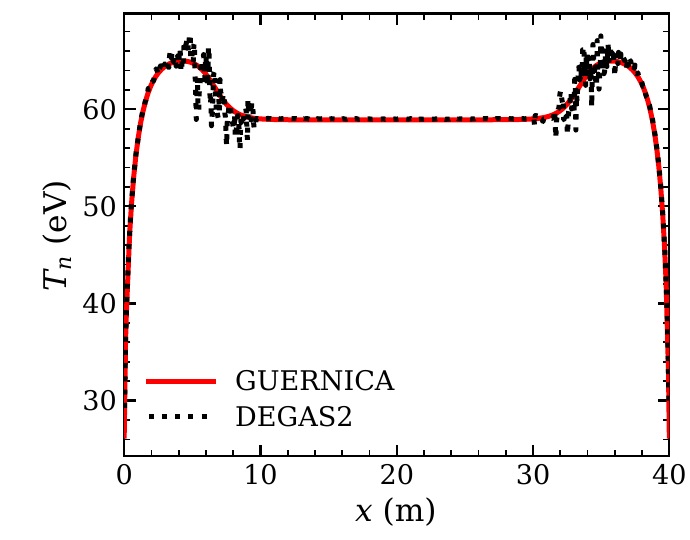}
\caption{Low density, neutral temperature}
\end{subfigure}\\[4pt]
\begin{subfigure}[t]{0.48\textwidth}\centering
\includegraphics[width=\textwidth]{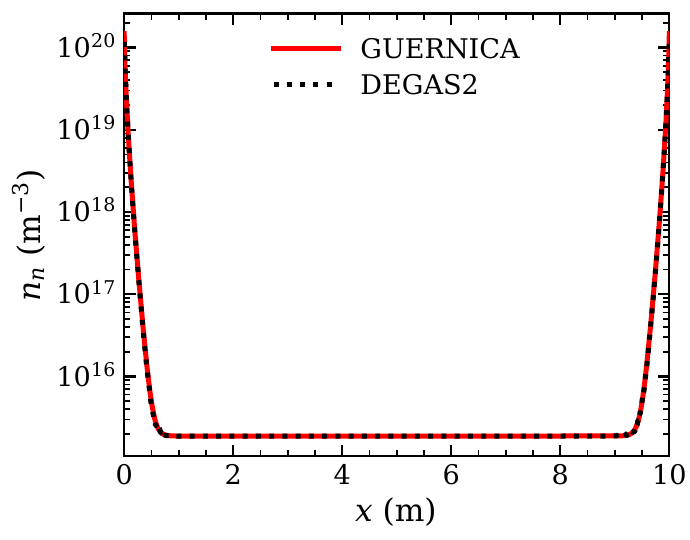}
\caption{High density, neutral density}
\end{subfigure}\hfill
\begin{subfigure}[t]{0.48\textwidth}\centering
\includegraphics[width=\textwidth]{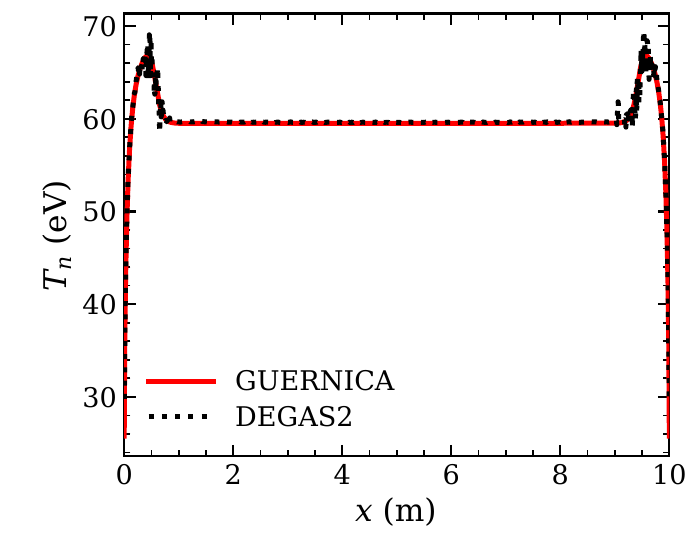}
\caption{High density, neutral temperature}
\end{subfigure}
\caption{Neutral density and temperature from GUERNICA and DEGAS2 for the (a, b) low-density and (c, d) high-density benchmarks with only charge exchange enabled.}
\label{fig:degas2_cx}
\end{figure}

Figure~\ref{fig:degas2} compares the neutral density and temperature from the two codes. The agreement is excellent throughout the domain. Away from the walls the two solutions agree within the statistical noise of DEGAS2, while close to the walls, where the DEGAS2 results are more precise, GUERNICA gives a density at most 1.6\% higher and a temperature within 1\% of DEGAS2. Repeating the calculations with coarser meshes, fewer velocity points and a larger velocity domain changed the solution by less than 0.6\% in every case. Evaluating the charge exchange convolution in single precision changed the high-density solution by $6\times10^{-7}$ while making each iteration 4.2 times faster.

We also repeated the benchmark with ionization only and with charge exchange only, using the same meshes and velocity grids. As seen in Figs.~\ref{fig:degas2_iz} and~\ref{fig:degas2_cx}, the agreement with DEGAS2 is as good as for the full model, showing that the two reactions are reproduced individually as well. In all cases the GUERNICA results are smooth and free of statistical noise.

\section{MAST-U demonstration}
\label{sec:MAST}

\begin{figure}[!htbp]
\centering
\begin{subfigure}[t]{0.32\textwidth}\centering
\includegraphics[width=\textwidth]{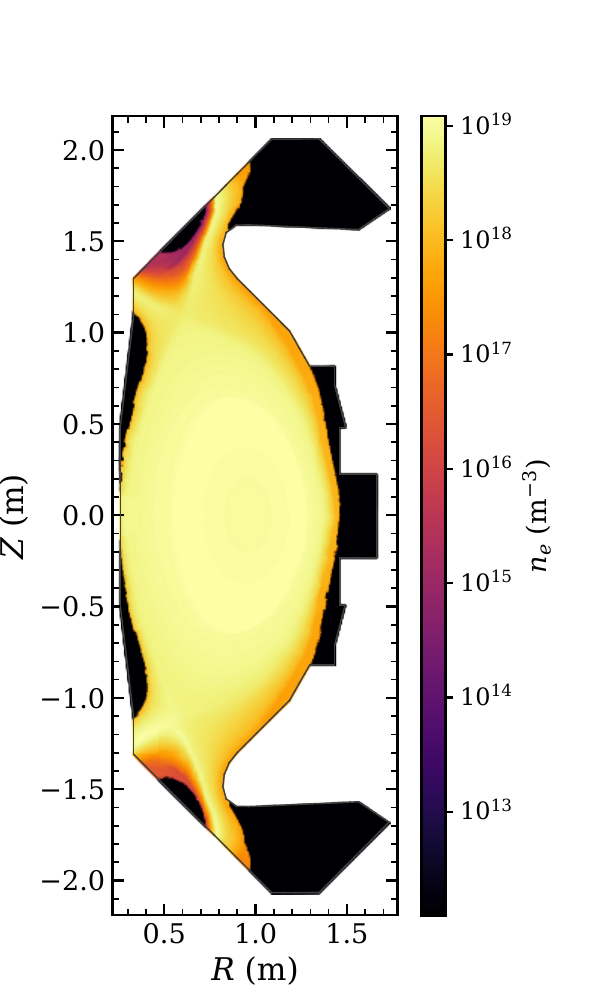}
\caption{Plasma density}
\end{subfigure}\hfill
\begin{subfigure}[t]{0.32\textwidth}\centering
\includegraphics[width=\textwidth]{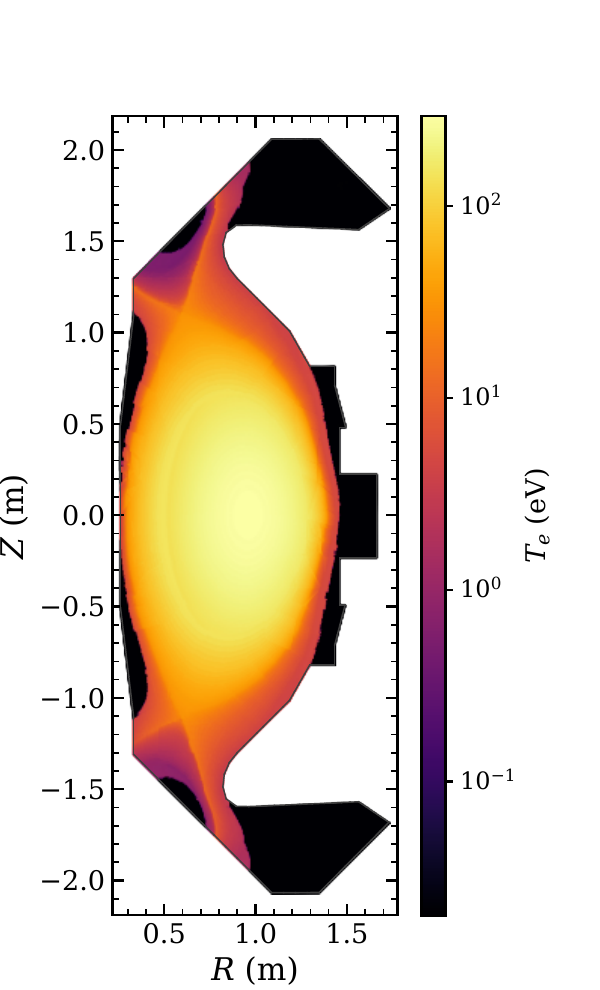}
\caption{Plasma temperature}
\end{subfigure}\hfill
\begin{subfigure}[t]{0.32\textwidth}\centering
\includegraphics[width=\textwidth]{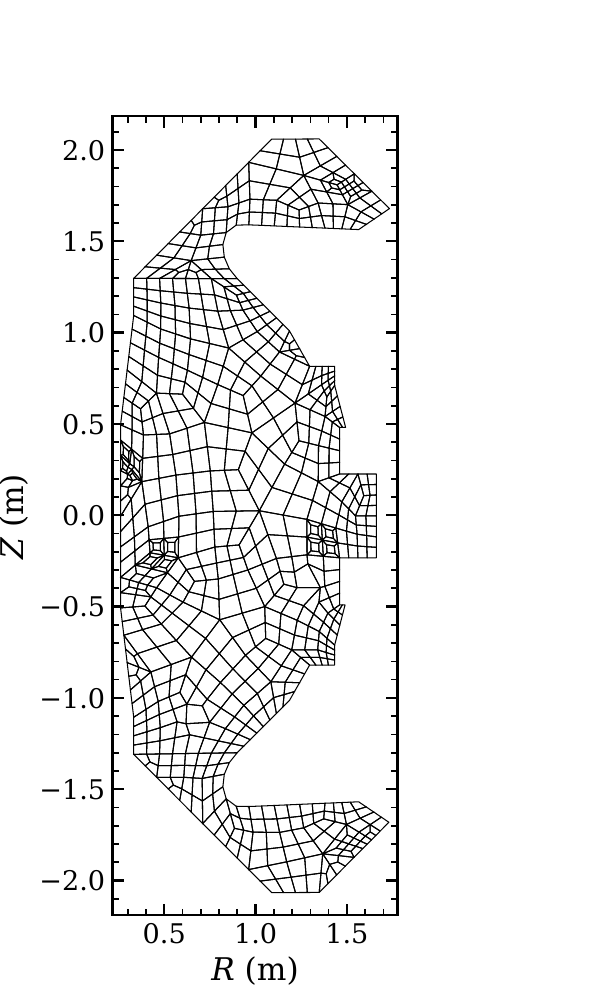}
\caption{GUERNICA mesh}
\end{subfigure}
\caption{Background plasma and mesh for the MAST-U demonstration. Plasma density (a) and temperature (b) come from the respective electron profiles from a SOLPS-ITER simulation with the ion fields set equal~\cite{moulton2024}.}
\label{fig:mastu_background}
\end{figure}

We present a preliminary demonstration of GUERNICA on a two-dimensional MAST-U cross section with experimentally relevant geometry and plasma conditions. The cross section is treated as a planar Cartesian domain that is uniform in the out-of-plane (toroidal) direction, so no cylindrical geometry terms are included and integrated quantities are given per meter in that direction. We took the background plasma from a SOLPS-ITER simulation of a MAST-U double-null, conventional divertor discharge from shot 45470 at 0.42~s~\cite{moulton2024}, an ohmically heated L-mode plasma with toroidal field $B_t = 0.5$~T and plasma current $I_p = 600$~kA.\footnote{SimDB alias 45470/Dec1623/seq-3, per Table A1 of~\cite{moulton2024}.} For simplicity, we set the ion density and temperature equal to their electron counterparts, shown in Fig.~\ref{fig:mastu_background}, and took the ion flow velocity from the same simulation.

The configuration-space domain extends beyond the plasma region to the first wall and contains 709 quadrilateral elements generated with Gmsh, as shown in Fig.~\ref{fig:mastu_background}c, with certain elements locally refined to keep the neutral density and temperature positive. We interpolated the plasma parameters from the B2 grid onto the GUERNICA mesh where they overlapped and used a background plasma of $n_e = 1\times10^8$~m$^{-3}$ and $T_e = 0.02$~eV with no flow for all regions outside the B2 grid, as is standard in SOLPS-ITER simulations. We used linear DG elements with $(N_{vx}, N_{vy}, N_{vz}) = (108, 108, 26)$ discrete velocity points spanning $[-4.7, 4.7]\times10^5$~m/s in each direction. At the walls, we used the recycling condition of Sec.~\ref{subsec:BCNum} with a wall temperature of 2~eV, so that each boundary face emits a neutral flux equal to the flux of the background ion flow onto it.

\begin{figure}[!htbp]
\centering
\includegraphics[width=0.48\textwidth]{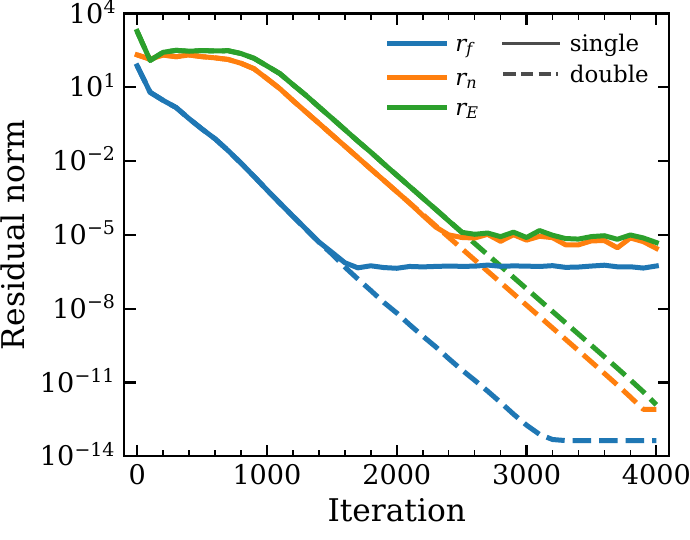}
\caption{Residual convergence for the MAST-U case over the steady-state pseudo-time iterations. The global distribution function residual, $r_f$, element-local density residual, $r_n$, and element-local energy residual, $r_E$, each fall by about eight orders of magnitude before leveling off at the round-off level of the single-precision charge exchange convolution (solid lines); the same case with double-precision charge exchange continues to $4\times10^{-14}$--$10^{-12}$ (dashed lines).}
\label{fig:mastu_residuals}
\end{figure}

We ran GUERNICA in steady-state mode with ionization, charge exchange and recombination active, with $\Delta\tau_0 = 0.05$ and the single-precision charge exchange convolution. Figure~\ref{fig:mastu_residuals} shows the convergence of $r_f$, $r_n$ and $r_E$ over 4000 pseudo-time iterations. All three decrease by about eight orders of magnitude before leveling off near $5\times10^{-7}$, $6\times10^{-6}$ and $8\times10^{-6}$. This level is set by round-off in the single-precision convolution. Repeating the calculation with double precision reduces the residuals to $4\times10^{-14}$--$10^{-12}$, as shown by the dashed lines in Fig.~\ref{fig:mastu_residuals}, and changes the neutral density by $2\times10^{-7}$ in the relative $L^2$ norm.

The 4000 iterations took 86~min on a single NVIDIA RTX A5000 for $8.6\times10^8$ phase-space degrees of freedom, with a peak memory of 18.8~GiB of the card's 24~GiB, against 6.0~h in double precision.

\begin{figure}[!htbp]
\centering
\begin{subfigure}[t]{0.32\textwidth}\centering
\includegraphics[width=\textwidth]{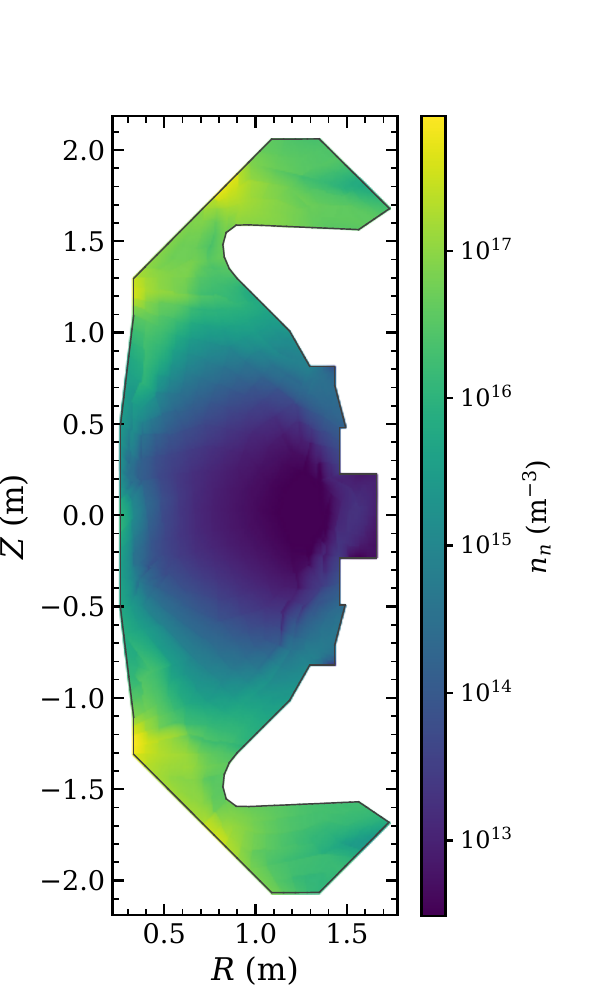}
\caption{Neutral density}
\end{subfigure}\hfill
\begin{subfigure}[t]{0.32\textwidth}\centering
\includegraphics[width=\textwidth]{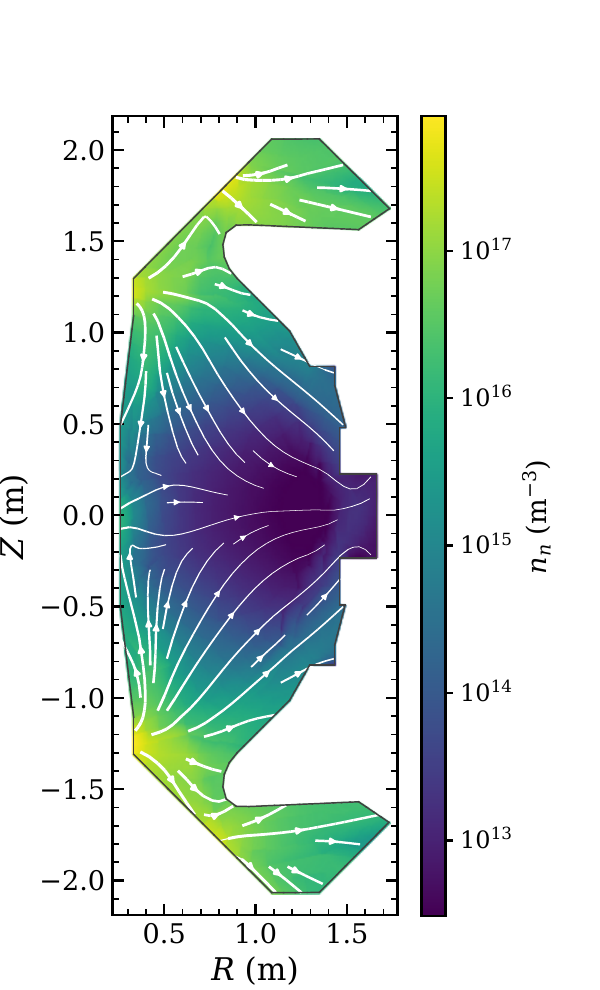}
\caption{Neutral flux}
\end{subfigure}\hfill
\begin{subfigure}[t]{0.32\textwidth}\centering
\includegraphics[width=\textwidth]{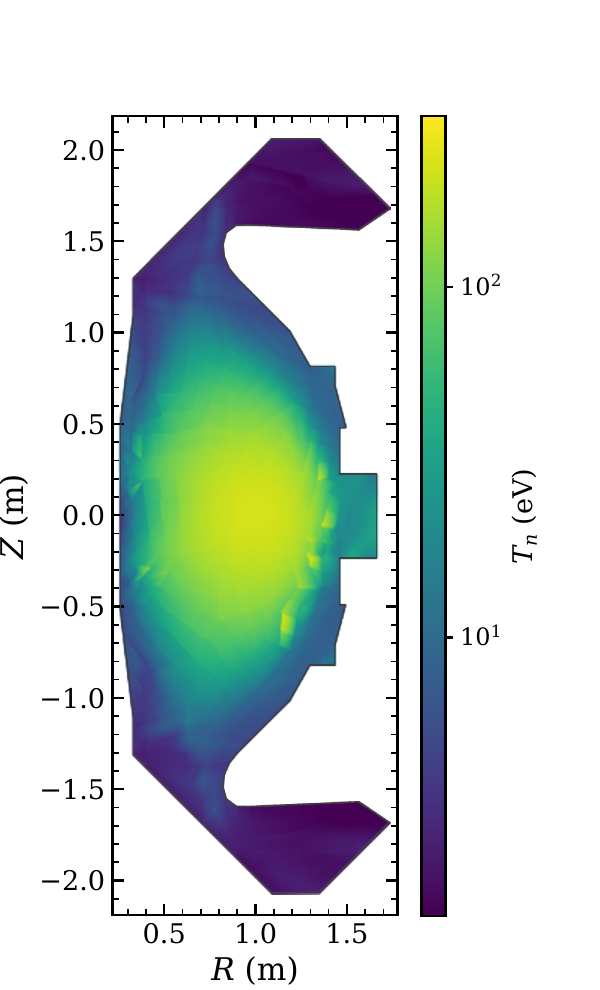}
\caption{Neutral temperature}
\end{subfigure}
\caption{Steady-state neutral solution for the MAST-U case.}
\label{fig:mastu_solution}
\end{figure}

Figure~\ref{fig:mastu_solution} shows the converged neutral density, flux and temperature. As expected, the neutral density is largest at the strike points where recycling takes place, reaching $8\times10^{17}$~m$^{-3}$ at the lower inner strike point, followed by the divertor chambers where no plasma is present, and drops off towards the scrape-off layer and core. Figure~\ref{fig:mastu_solution}b shows streamlines of the steady-state neutral flux on top of $n_n$, showing the transport of neutrals from the strike points through the domain until they reach a wall where they are absorbed. Reflection of neutrals at the walls, which would return part of this flux to the domain, is left for future work.

The neutral temperature in Fig.~\ref{fig:mastu_solution}c rises from a few eV in the divertor chambers and near the strike points to above 100~eV in the core, where the few neutrals that penetrate have undergone charge exchange with hot ions. In the chambers the neutral temperature is close to the 2~eV wall temperature, but the local values there are sensitive to the out-of-plane velocity resolution discussed below, so the temperature of the cold neutrals should be read qualitatively.

\begin{figure}[!htbp]
\centering
\begin{subfigure}[t]{0.48\textwidth}\centering
\includegraphics[width=\textwidth]{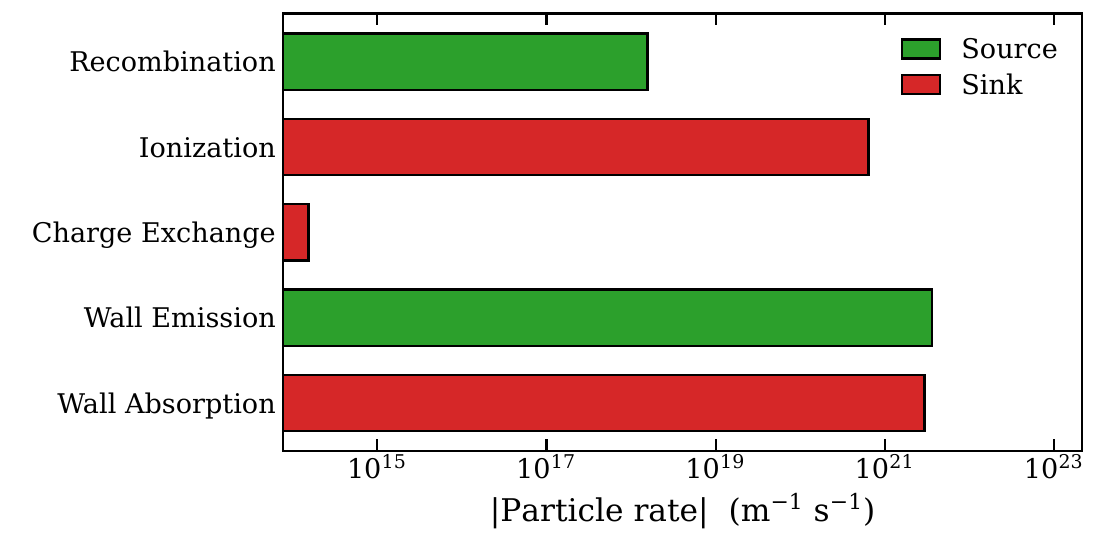}
\caption{Neutral particle balance}
\end{subfigure}\hfill
\begin{subfigure}[t]{0.48\textwidth}\centering
\includegraphics[width=\textwidth]{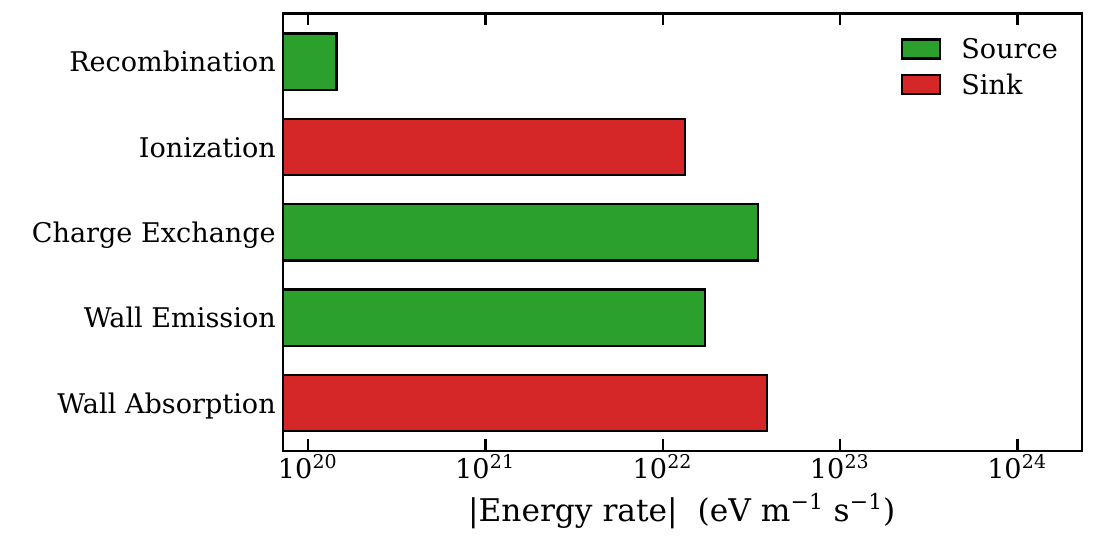}
\caption{Neutral energy balance}
\end{subfigure}
\caption{Global integrated particle and energy balance for the collision operators, recombination source and the walls.}
\label{fig:mastu_balance}
\end{figure}

\begin{figure}[!htbp]
\centering
\begin{subfigure}[t]{0.32\textwidth}\centering
\includegraphics[width=\textwidth]{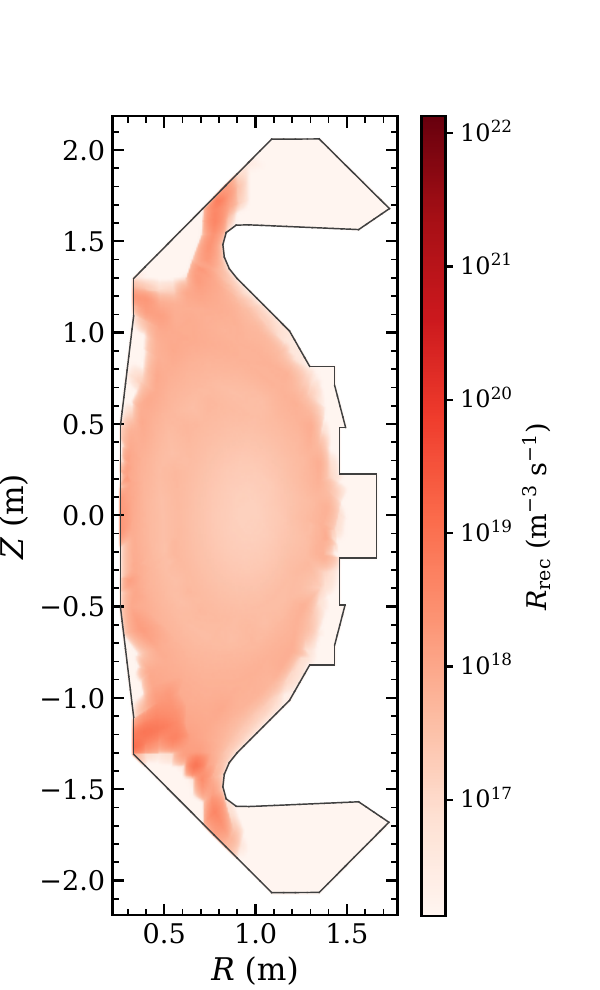}
\caption{$R_{rec}$}
\end{subfigure}\hfill
\begin{subfigure}[t]{0.32\textwidth}\centering
\includegraphics[width=\textwidth]{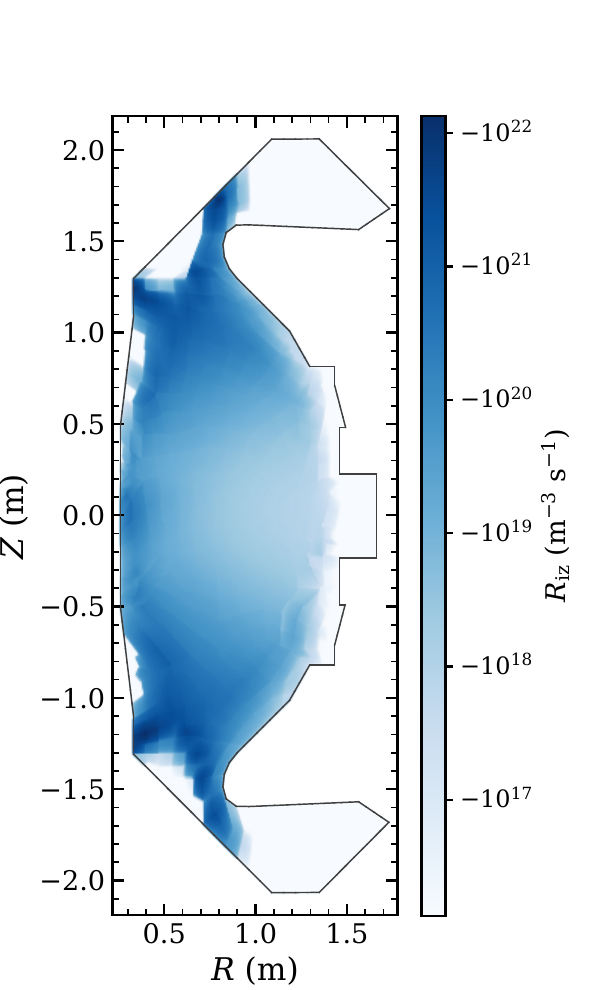}
\caption{$R_{iz}$}
\end{subfigure}\hfill
\begin{subfigure}[t]{0.32\textwidth}\centering
\includegraphics[width=\textwidth]{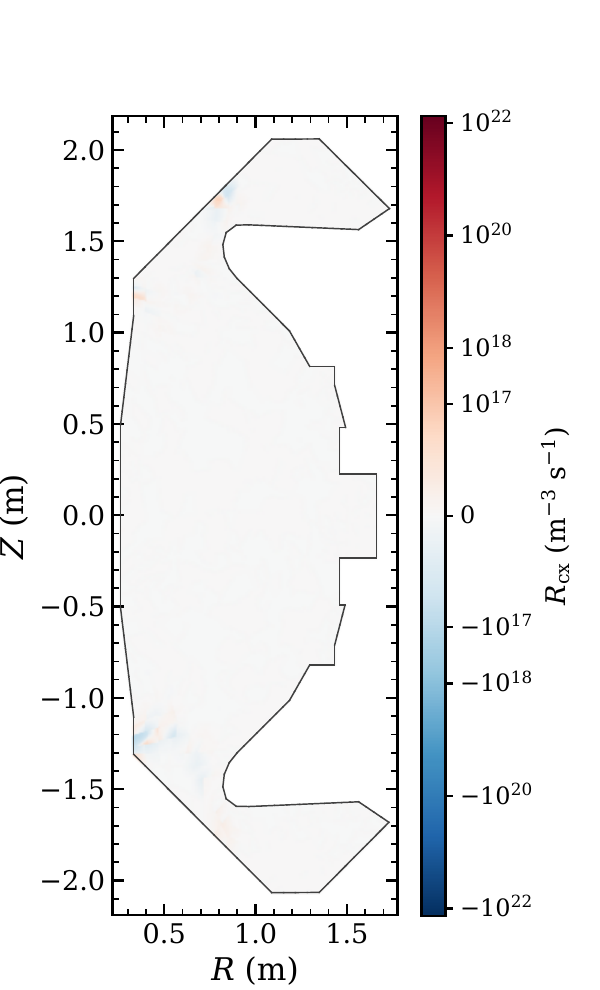}
\caption{$R_{cx}$}
\end{subfigure}\\[4pt]
\begin{subfigure}[t]{0.32\textwidth}\centering
\includegraphics[width=\textwidth]{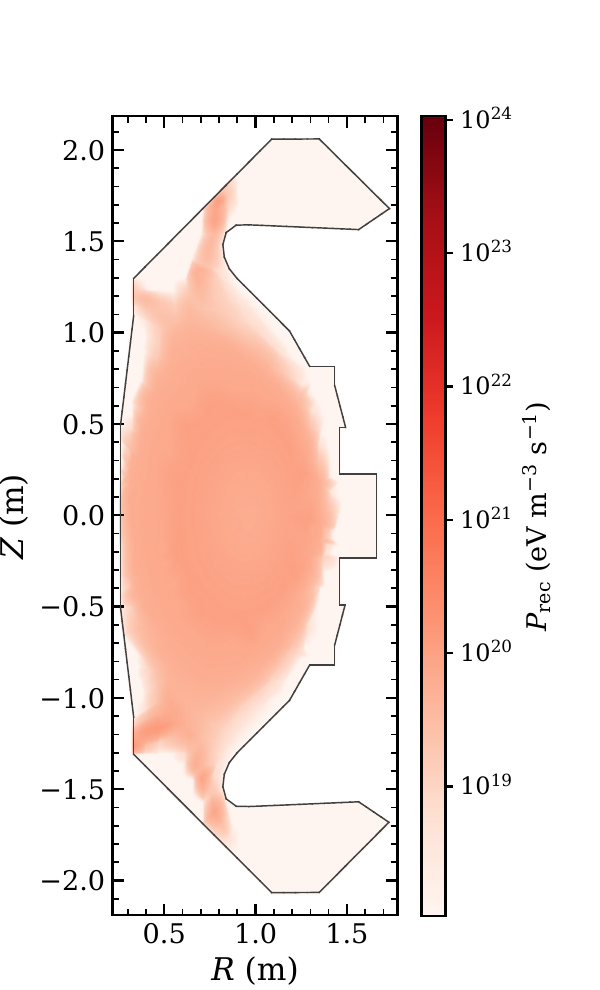}
\caption{$P_{rec}$}
\end{subfigure}\hfill
\begin{subfigure}[t]{0.32\textwidth}\centering
\includegraphics[width=\textwidth]{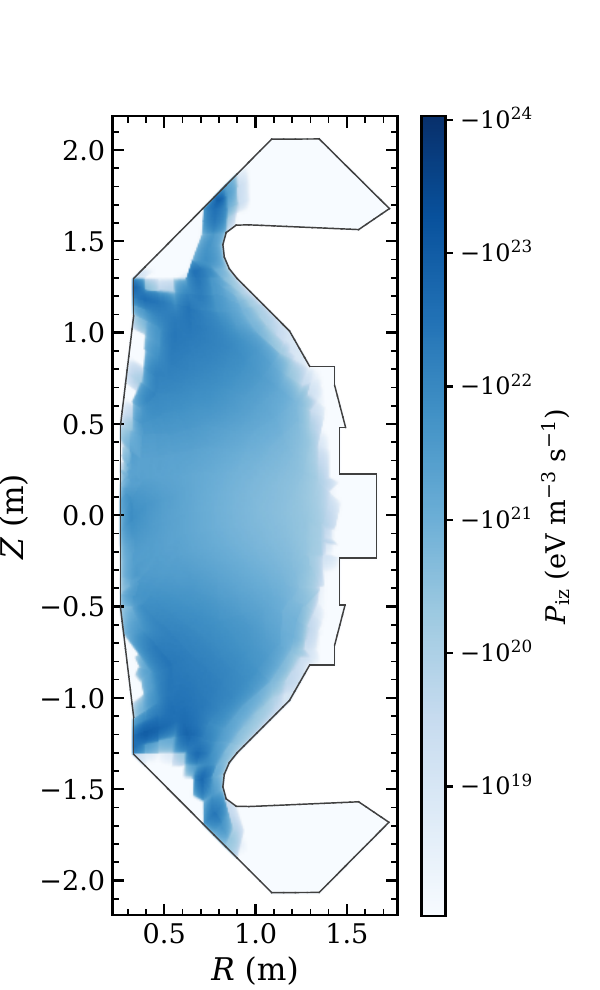}
\caption{$P_{iz}$}
\end{subfigure}\hfill
\begin{subfigure}[t]{0.32\textwidth}\centering
\includegraphics[width=\textwidth]{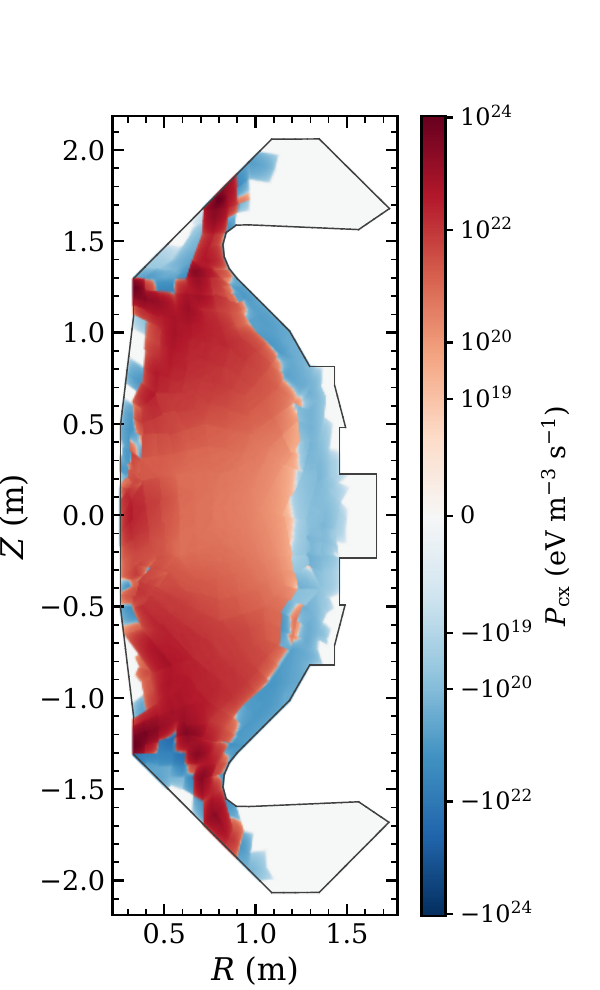}
\caption{$P_{cx}$}
\end{subfigure}
\caption{Steady-state particle rates $R$ (top row) and energy rates $P$ (bottom row) for recombination, ionization and charge exchange.}
\label{fig:mastu_rates}
\end{figure}

To characterize the plasma--neutral interactions, Fig.~\ref{fig:mastu_balance} shows the global integrated neutral particle and energy balance, and Fig.~\ref{fig:mastu_rates} shows the steady-state particle rates, $R_{rec}$, $R_{iz}$ and $R_{cx}$, and energy rates, $P_{rec}$, $P_{iz}$ and $P_{cx}$, of recombination, ionization and charge exchange. Each is the density or energy moment of the corresponding term of Eq.~\ref{eq:Gov}, so positive values are sources of neutrals or of neutral energy. The integrated particle and energy balances close to a relative $3\times10^{-9}$ or better. Recombination plays a small role as both a particle and energy source, with an integrated particle source of $1.6\times10^{18}$~m$^{-1}$s$^{-1}$ against $3.6\times10^{21}$~m$^{-1}$s$^{-1}$ emitted by the walls, only playing an appreciable role near the strike points and along the divertor legs, where the plasma density is high and temperature is low. Three-body recombination, which would be largest in these regions, is not yet included, as noted in Sec.~\ref{subsec:ColGov}. Ionization plays a key role as a sink of both neutral particles and energy, limiting neutrals' ability to penetrate deeply into the core in Fig.~\ref{fig:mastu_solution}a. It removes 18\% of the emitted neutrals, and the remaining 82\% are absorbed at the walls. Charge exchange is the largest source in the energy balance, transferring energy from hot ions to the neutrals over almost the whole domain, with thin layers near the walls, most visibly on the outboard side, where it cools hot neutrals. Since charge exchange conserves particle number, $R_{cx}$ does not play a role in particle balance, with the small values in Fig.~\ref{fig:mastu_rates}c, which integrate to $4\times10^{-8}$ of the wall emission, coming from single-precision round-off. In the double-precision run they fall to $3\times10^{-16}$.

Faint streaks originating at the strike points are visible in the density of Fig.~\ref{fig:mastu_solution}a. These are the ray effect examined in Sec.~\ref{sec:transport}, caused by an under-resolved velocity discretization. The in-plane velocity grid has about 1.6 points per thermal speed of the 2~eV recycled neutrals, between the $16^2$ and $32^2$ grids of the emitting wall test, where the ray effect is 15\% and 3\%.\footnote{Similar streak-like features appear in SOLPS-ITER simulations with too few EIRENE particles, although their origin is statistical rather than discretization-related.} The out-of-plane direction is coarser still, with 26 points 38~km/s apart against a thermal speed of 14~km/s for the recycled neutrals. As a result the emitted neutrals carry a mean energy of 4.8~eV instead of the 4~eV of a 2~eV Maxwellian flux, so the energy rates and the temperature of the cold neutrals are qualitative. Resolving these neutrals in the out-of-plane direction would require about 46 points, which exceeds the memory of a single GPU at this in-plane resolution. Using a spherical velocity grid instead of a uniform Cartesian velocity grid would ameliorate this issue; however, it would require overhauling the FFT-based evaluation of the charge exchange operator.

\section{Summary and future work}
\label{sec:Conc}

This paper introduced GUERNICA, a deterministic continuum kinetic code for the transport of atomic hydrogen neutrals in the edge and divertor regions of fusion plasmas. It combines the discrete velocity method in velocity space with a matrix-free, GPU-accelerated discontinuous Galerkin discretization in configuration space built on MFEM, and includes charge exchange, ionization, BGK neutral--neutral collisions and a recombination source.

Tests of the individual operators showed that the transient integrator is third order, the BGK operator conserves neutral mass, momentum and energy to near machine precision while increasing the discrete entropy, and the charge exchange rate converges at fourth order in the velocity spacing. Against exact solutions of the transport problem, the DG discretization converges at the expected $O(\Delta x^{p+1})$ rate for $p = 1$ to 4, including on randomly perturbed meshes, and the diffuse wall boundary condition converges at the same rate.

The same tests showed that the velocity quadrature converges exponentially, twice as fast with an even number of points as with an odd number, and quantified the ray effect that dominates the error in problems with localized sources at practical velocity resolutions.  The steady-state solver is limited by a transport limit on the pseudo-time step at low collisionality, which decreases with the polynomial degree, and by a collisional limit at high collisionality, which does not, so that higher orders are more accurate per node but require more iterations in weakly collisional regions. With the FFT evaluation, the cost of the charge exchange operator grows as $O(N_v\log N_v)$ and is three orders of magnitude below that of a direct evaluation at $48^3$ velocity points with $O(N_v^2)$ scaling, while advection, ionization and BGK scale linearly. Finally, GUERNICA reproduces the DEGAS2 neutral density and temperature for the one-dimensional benchmarks of Bernard et al.\ within the Monte Carlo noise away from the walls and to within 1.6\% near them, without the statistical noise inherent to Monte Carlo sampling.

A two-dimensional demonstration on a realistic, unstructured MAST-U double-null geometry, using a SOLPS-ITER background plasma with ionization, charge exchange and recombination, converged to the round-off level of the single-precision charge exchange convolution in 86~min on a single GPU. The solution closes the global particle and energy balances to a relative $3\times10^{-9}$ or better and resolves the spatial structure of each reaction rate without statistical noise. The limited velocity resolution of the cold recycled neutrals, however, produces a visible ray effect. Together these results demonstrate the feasibility of noise-free continuum neutral transport in realistic geometry on a single GPU.

Several extensions are planned. First, the uniform Cartesian velocity grid ties the required number of velocity points to the ratio of the highest to lowest temperature to be resolved; in the MAST-U case it limits the velocity resolution of the cold recycled neutrals and produces the ray effect seen in the density. To relax this constraint, we are exploring a spherical $(v_r, v_\theta, v_\phi)$ velocity discretization, which will require reformulating the FFT-based charge exchange convolution. Second, we are developing a moment-preserving pseudo-time treatment to run the BGK neutral--neutral operator in steady-state mode. Third, we will extend the physics model to three-body recombination, reflection of neutrals at the walls and molecular hydrogen, which plays an important role in detachment. Finally, we will extend the present single-GPU implementation to multi-GPU, distributed-memory execution as necessary to reach the resolutions required for reactor-scale geometries such as ITER.

The ultimate goal is the coupling of GUERNICA to a plasma edge fluid or gyrokinetic solver, replacing the static background plasma used throughout this paper with a self-consistently evolving one. The noise-free, deterministic nature of GUERNICA's solutions is intended to make this iterative coupling more robust than is presently possible with Monte Carlo neutral models, particularly at the high recycling coefficients characteristic of detached divertor operation.

\section*{CRediT authorship contribution statement}
\textbf{Jack Gabriel:} Conceptualization, Methodology, Software, Validation, Formal analysis, Investigation, Data curation, Visualization, Writing -- original draft, Writing -- review and editing.

\textbf{Saskia Mordijck:} Conceptualization, Methodology, Supervision, Funding acquisition, Resources, Project administration, Writing -- review and editing.

\section*{Acknowledgments}
The authors thank Ekin \"{O}zt\"{u}rk for providing the MAST-U SOLPS-ITER data and explaining how it is structured, and George Wilkie for supplying the DEGAS2 results for the benchmark of Bernard et al.~\cite{bernard2022}. We also thank Tess Bernard and Jimmy Juno for helpful discussions of the benchmark.

\section*{Funding} This material is based upon work supported by the U.S. Department of Energy, Office of Science, Office of Fusion Energy Sciences under Awards DE-SC0007880, DE-SC0021306 and DE-SC0023372

\appendix
\section{Exact solutions of the periodic source test}
\label{app:fourier}

With charge exchange and neutral--neutral collisions turned off and a constant ionization rate $\nu$, the steady kinetic equation for the periodic source test of Sec.~\ref{sec:transport} is
\begin{equation}
v_x\frac{\partial f}{\partial x} + \nu f = \left[c_0 + c_1\cos kx\right] M(\mathbf{v}),
\label{eq:app_kinetic}
\end{equation}
on a periodic domain of one wavelength, $kL_x = 2\pi$, where $M(\mathbf{v})$ is a unit Maxwellian at rest with temperature $T$ and the product over the three velocity directions of the one-dimensional factor $M(v) = (2\pi T)^{-1/2}\exp(-v^2/2T)$. The velocity appears in Eq.~\ref{eq:app_kinetic} only as a parameter, so each velocity satisfies a separate first-order linear equation in $x$ with constant coefficients.

The constant part of the source is balanced by a uniform solution, $f_0 = (c_0/\nu)M(\mathbf{v})$. For the oscillating part we write $\cos kx = \mathrm{Re}\,e^{ikx}$ and seek $f_1 = \mathrm{Re}\,[F(\mathbf{v})e^{ikx}]$, for which Eq.~\ref{eq:app_kinetic} gives $(\nu + ikv_x)F = c_1 M(\mathbf{v})$. Taking the real part,
\begin{equation}
f_1(x, \mathbf{v}) = c_1 M(\mathbf{v})\, \frac{\nu\cos kx + kv_x\sin kx}{\nu^2 + k^2v_x^2}.
\label{eq:app_f1}
\end{equation}
The homogeneous solutions $\exp(-\nu x/v_x)$ are not periodic, so $f = f_0 + f_1$ is the only steady solution.

Integrating over velocity, the Maxwellian integrates to one, the $\sin kx$ term vanishes because its integrand is odd in $v_x$, and the integrals over $v_y$ and $v_z$ give one because the remaining integrand depends only on $v_x$. The density is therefore
\begin{equation}
n(x) = \frac{c_0}{\nu} + c_1 A\cos kx, \qquad
A = \int_{-\infty}^{\infty} M(v_x)\,\frac{\nu}{\nu^2 + k^2v_x^2}\,dv_x.
\label{eq:app_A}
\end{equation}

For the continuous problem, the substitution $t = v_x/\sqrt{2T}$ gives
\begin{equation}
A = \frac{1}{\sqrt{\pi}}\,\frac{\nu}{2Tk^2}\int_{-\infty}^{\infty} \frac{e^{-t^2}}{t^2 + a^2}\,dt, \qquad 
a = \frac{\nu}{k\sqrt{2T}},
\end{equation}
and the standard integral $\int_{-\infty}^{\infty} e^{-t^2}/(t^2 + a^2)\,dt = (\pi/a)\,e^{a^2}\operatorname{erfc}(a)$ yields
\begin{equation}
A = \sqrt{\frac{\pi}{2}}\,\frac{1}{k\sqrt{T}}\operatorname{erfcx}\!\left(\frac{\nu}{k\sqrt{2T}}\right),
\label{eq:app_A_cont}
\end{equation}
where $\operatorname{erfcx}(z) = e^{z^2}\operatorname{erfc}(z)$ is the scaled complementary error function.

GUERNICA solves the same problem with the velocity integrals replaced by the quadrature of Sec.~\ref{subsec:DVM}. Each discrete velocity still satisfies Eq.~\ref{eq:app_kinetic}, so Eq.~\ref{eq:app_f1} holds at every velocity point. The discrete Maxwellian is normalized so that its quadrature sum in each direction is exactly one, which makes the transverse sums equal to one, and the $v_x$ points are symmetric about zero, so the $\sin kx$ term cancels exactly. The discrete-velocity density therefore has the form of Eq.~\ref{eq:app_A} with the amplitude
\begin{equation}
A_d = \sum_j \hat{w}_j\,\frac{\nu}{\nu^2 + k^2v_{x,j}^2}, \qquad
\hat{w}_j = \frac{w_j M(v_{x,j})}{\sum_i w_i M(v_{x,i})},
\label{eq:app_A_disc}
\end{equation}
where $w_j$ are the one-dimensional quadrature weights and $v_{x,j}$ the velocity points.

\newpage
\bibliographystyle{elsarticle-num}
\bibliography{cas-refs}

\end{document}